\documentclass[preprint,sort&compress,11pt]{elsarticle}

\usepackage{graphicx}
\usepackage{amsmath}
\usepackage{natbib}
\usepackage{amssymb}
\usepackage{amsthm}
\usepackage{lineno}
\usepackage{subfig}
\usepackage{enumerate}
\usepackage{fullpage}
\usepackage{algorithm}
\usepackage{xcolor}
\usepackage{multirow}
\usepackage{bbm}
\usepackage{fullpage}
\usepackage{url}
\usepackage{listings}
\usepackage[colorlinks=true]{hyperref}

\usepackage{color,soul}
\usepackage{enumitem}
\usepackage{mathtools}

\biboptions{sort,compress} 
\usepackage{array}
\newcolumntype{P}[1]{>{\centering\arraybackslash}m{#1}}
\newcommand{\RN}[1]{\textup{\uppercase\expandafter{\romannumeral#1}}}

\usepackage{dashrule}

\usepackage{changes}
\definechangesauthor[name={Reviewer 1}, color = blue]{R1}
\definechangesauthor[name={Reviewer 2}, color = purple]{R2}
\definechangesauthor[name={All reviewers}, color = brown]{All}
\definechangesauthor[name={JX Wang}, color = olive]{Author}
\definechangesauthor[name={Heng Xiao}, color = orange]{hx}

\journal{Physical Review Fluids}

\begin{document}

\begin{frontmatter}
	
  

 \title{A Physics Informed Machine Learning Approach for Reconstructing Reynolds Stress Modeling Discrepancies Based on DNS Data}

	
	
  \author{Jian-Xun Wang}
  \author{Jin-Long Wu}
  \author{Heng Xiao\corref{corxh}}
  \cortext[corxh]{Corresponding author. Tel: +1 540 231 0926}
  \ead{hengxiao@vt.edu}
  
  \address[vt]{Department of Aerospace and Ocean Engineering, Virginia Tech,
    Blacksburg, VA 24060, United States}

  \begin{abstract}
    Turbulence modeling is a critical component in numerical simulations of industrial flows based
    on Reynolds-averaged Navier-Stokes (RANS) equations. However, after decades of efforts in the
    turbulence modeling community, universally applicable RANS models with predictive capabilities
    are still lacking. {Large discrepancies in the RANS-modeled Reynolds stresses are the main source that limits the predictive accuracy of RANS models. Identifying these discrepancies is of significance to possibly improve the RANS modeling. In this work, we propose a data-driven, physics-informed machine learning approach for reconstructing discrepancies in RANS modeled Reynolds stresses.} The discrepancies are formulated as functions of the mean flow features.  By using a modern machine learning technique based on random forests, {the discrepancy functions are trained by existing DNS databases and then used to predict Reynolds stress discrepancies in different flows where data are not available.} 
	The proposed method is evaluated by two classes of 
    flows: (1) fully developed turbulent flows in a square duct at various Reynolds numbers 
    and (2) flows with massive separations. In separated flows, two training flow scenarios 
    of increasing difficulties are considered: (1) the flow in the same periodic hills geometry 
	yet at a lower Reynolds number, and (2) the flow in a different
    hill geometry with a similar recirculation zone. Excellent predictive performances were
    observed in both scenarios, demonstrating the merits of the proposed method.
  \end{abstract}
	
\end{frontmatter}


\section{Introduction}
\label{sec:intro}

\subsection{RANS Models as Workhorse Tool in Industrial CFD}
Computational fluid dynamics (CFD) simulations have been widely used in aerospace, mechanical, and
chemical industries to support engineering design, analysis, and optimization.  Two decades ago when
Large Eddy Simulations (LES) started gaining popularity with the increasing availability of
computational resources, it was widely expected that LES would gradually displace and eventually
replace Reynolds-Averaged Navier-Stokes (RANS) equations in industrial Computational Fluid Dynamics
(CFD) work-flows for decades to come.  In the past two decades, however, while LES-based methods
(including resolved LES, wall-modeled LES, and hybrid LES/RANS methods) did gain widespread
applications, and the earlier hope certainly did not diminish, the predicted time when these methods
would replace RANS has been significantly delayed. This observation is particularly relevant in
light of the recent discussions on the ending of the ``{Moore's Law} era'' with transistor sizes
approaching their theoretical lower limit~\cite{waldrop2016chips,kumar2015fundamental}.  RANS
solvers, particularly those based on standard eddy viscosity models (e.g.,
$k$--$\varepsilon$~\cite{launder1974application},
$k$--$\omega$~\cite{wilcox1988reassessment,wilcox2006turbulence}, S--A ~\cite{spalart1992one}, 
and k--$\omega$--SST~\cite{menter1994two}, are still and will remain the dominant tool for industrial
CFD in the near future. This is likely to be true even in mission critical applications such as
aircraft design.  Interestingly, even the advanced RANS models such as Reynolds stress transport
models~\cite{launder1975progress} and Explicit Algebraic Reynolds stress
models~\cite{wallin2000explicit} have not seen much development in the past few decades. These
advanced models are computationally more expensive and less robust compared to the standard eddy
viscosity RANS models. As such, it is still practically important to further develop the standard
RANS models for industrial CFD applications. However, improving the predictive capabilities of these
models is critical yet technically challenging.

\subsection{Progress and Challenges in Data-Driven Turbulence Modeling} 

While traditional development of turbulence models has focused on incorporating more physics to
improve predictive capabilities, an alternative approach is to utilize data.  In the past few years,
a number of data-driven approaches have been proposed. Researchers have investigated the use of both
offline data (i.e., existing DNS data for flows different from that to be
predicted~\cite{dow11quanti,parish2016paradigm,ling2015evaluation}) and online data (streamed
monitoring data from the flow to be predicted~\cite{xiao-mfu,wang2016data,iungo2015data}).  Dow and
Wang~\cite{dow11quanti} used Direct Numerical Simulation (DNS) data from a plane channel flow to
infer the full-field discrepancy in the turbulent viscosity $\nu_t$ modeled by the $k$--$\omega$
model.  To predict flows in channels with wavy boundaries, they modeled the (log-)discrepancies of
$\nu_t$ in the new flows as Gaussian random fields, with the discrepancy field inferred above as
mean.  Duraisamy and co-workers~\cite{parish2016paradigm,singh16using} introduced a full-field
multiplicative discrepancy term $\beta$ into the production term of the transport equations of
turbulent quantities (e.g., $\tilde{\nu}_t$ in the SA model and $\omega$ in the $k$--$\omega$
models).  They used DNS data to calibrate and infer uncertainties in the $\beta$ term. It is
expected that the inferred discrepancy field can provide valuable insights to the development of
turbulence model and can be used to improve RANS predictions in similar flows. Xiao et
al.~\cite{xiao-mfu} used sparse velocity measurements (online data) to infer the full-field
discrepancies $\Delta \tau_\alpha$ in the RANS-predicted Reynolds stress tensors, or more precisely
the physical projections thereof (turbulent kinetic energy, anisotropy, and
orientations). Throughout this paper it is understood that $\tau_\alpha$ indicates the physical
projections and not the individual components of the Reynolds stress tensor.  Good performance was
demonstrated on several canonical flows including flow past periodic hills, flow in a square
duct~\cite{xiao-mfu}, and flow past a wing--body junction~\cite{wu2016wing}.

All three approaches~\cite{dow11quanti,parish2016paradigm,xiao-mfu} discussed above can be
considered starting points toward the same destination: the capability of predictive turbulence
modeling by using standard RANS models in conjunction with offline data. To this end, the respective
discrepancies terms ($\Delta \log \nu_t$, $\beta$, and $\Delta \tau_\alpha$) are expected to be
extrapolated to similar yet different flows.  These contributions are all relatively recent and much
of the research is still on-going.  Duraisamy et al.~\cite{singh16using,duraisamy2015new}
performed \textit{a priori} studies to show the potential universality of their discrepancy term
$\beta$ among a class of similar flows, but their performances in \textit{a posteriori} tests, i.e.,
using the calibrated discrepancy in one flow to predict another flow, have yet to be demonstrated.
Dow and Wang~\cite{dow11quanti} extrapolated the logarithmic discrepancies $\Delta \log \nu_t$
calibrated in the plane channel flow to flows in channel with slightly wavy walls, where velocity
predictions were made.  Similarly, further pursuing the approach of Xiao et al.~\cite{xiao-mfu}, Wu
et al.~\cite{wu2016bayesian} showed that the Reynolds stress discrepancy calibrated with sparse
velocity data can be extrapolated to flows at Reynolds number more than an order of magnitude higher
than that in the calibration case.  The extrapolated discrepancy has lead to markedly improved
predictions of velocities and other Quantities of Interest (QoIs), showing the potential of the
approach in enabling data-driven predictive turbulence modeling.  However, an intrinsic limitation
in the approach of Wu et al.~\cite{wu2016bayesian} is that they inferred the functions
$f^{(x)}_\alpha: \mathbf{x} \mapsto \Delta \tau_\alpha$, or simply denoted as $\Delta \tau_\alpha
(\mathbf{x})$, in the space of \emph{physical coordinates} $\mathbf{x}$. Therefore, strictly
speaking they only demonstrated that the discrepancy $\Delta \tau_\alpha$ can be extrapolated to
flows in the same geometry at the same location.  Consequently, their attempts of extrapolation to
the flow in a different geometry (e.g., from a square duct to a rectangular duct) encountered
less success.  The approach of Dow and Wang~\cite{dow11quanti} would share the same limitation since
they built Gaussian random fields indexed by the physical coordinates $\mathbf{x}$.

\subsection{Motivation of the Proposed Approach}
\label{sec:motivation}

A natural extension that overcomes the key limitation in the calibration--prediction approach of Wu
et al.~\cite{wu2016bayesian} is to build such functions in a space of well-chosen features
$\mathbf{q}$ instead of physical coordinates $\mathbf{x}$.  Despite its limitations, a key factor in
the success of the original approach is that the Reynolds stress discrepancies are formulated on its
projections such as the anisotropy parameters ($\xi$ and $\eta$) and orientation ($\varphi_i$) of
the Reynolds stresses and not directly on the individual components.  These projections are
normalized quantities~\cite{wu2016bayesian}. We shall retain this merits in the current approach and
thus use data to construct functions $\Delta \tau_\alpha(\mathbf{q})$ instead of $\Delta
\tau_\alpha(\mathbf{x})$. This extension would allow the calibrated discrepancies to be
extrapolated to a much wide range of flows. {In other words, the discrepancies of the 
RANS-predicted Reynolds stresses can be quantitatively explained by the mean flow physics.} Hence, these discrepancies are likely to be universal quantities that can be extrapolated 
from one flow to another, at least among different flows sharing the same characteristics (e.g., separation).  
As such, discrepancies in Reynolds stress projections are suitable targets to build functions for.


With the function targets identified, two challenges remain: (1) to identify a set of mean flow
features based on which the discrepancies functions $\Delta \tau_\alpha(\mathbf{q})$ can be
constructed and (2) to choose a suitable method for constructing such functions. Duraisamy and
co-workers~\cite{duraisamy2015new} identified several features and used neural network to construct
functions for the multiplicative discrepancy term. 
Ling and Templeton~\cite{ling2015evaluation} provided a richer and much more complete
set of features in their pioneering work, and they evaluated several machine learning algorithms to
predict point-based binary confidence indicators of RANS models~\cite{ling2015evaluation}. Ling et
al.~\cite{ling2016turbo} further used machine learning techniques to predict the Reynolds stress
anisotropy in jet-in-cross flows. Based on the success demonstrated by Ling and
co-workers~\cite{ling2015evaluation,ling2016turbo}, we will use machine learning to construct the
functions $\Delta \tau_\alpha(\mathbf{q})$ in the current work. Specifically, we will examine a
class of supervised machine learning techniques, where the objective of the learning is to build  a
statistical model from data and to make predictions on a response based on one or more
inputs~\cite{friedman2001elements}. This is in contrast to unsupervised learning, where no response
is used in the training or prediction, and the objective is to understand the relationship and
structure of the input data. Unsupervised learning will be explored as an alternative approach in
future works.

\subsection{Objective, Scope, and Vision of This Work}
{The objective of this contribution is to present an approach to predict Reynolds stress modeling discrepancies in new flows by utilizing data from flows with similar characteristics as the prediction flow.} This is achieved by training regression functions of Reynolds stress discrepancies with the DNS database from the training flows.

In light of the consensus in the turbulence modeling community that the Reynolds stresses are the
main source of model-form uncertainty in RANS
simulations~\cite{oliver2009uncertainty,wilcox2006turbulence,pope2000turbulent}, the current work
aims to improve the RANS modeled Reynolds stresses.  In multi-physics applications the QoIs might
well be the Reynolds stresses and/or quantities that directly depend thereon.  In these applications
the current work is significant by itself in that it would enable the use of standard RANS models in
conjunction with an offline database to provide accurate Reynolds stress predictions. 
Moreover, the improvement of Reynolds stresses enabled by the proposed method
is an important step towards a data-driven turbulence modeling framework. However, the Reynolds
stresses corrected by the constructed discrepancy function from DNS databases cannot necessarily 
guarantee to obtain improved mean flow fields. There are a number of challenges associated 
with propagating the improvement of Reynold stresses through RANS equation to the mean velocity field, which will be addressed in future works.

The rest of this paper is organized as follows. Section~2 introduces the components of the predictive
framework, including the choice of regression inputs and responses as well as the machine learning
technique used to build the regression function. Section~3 shows the numerical results to
demonstrate the merits of the proposed method. Further interpretation of the feature importance and
its implications to turbulence model development are discussed in Section~4. Finally, Section~5
concludes the paper.

\section{Methodology}
\label{sec:method}

\subsection{Problem Statement}
The overarching goal of the current and companion works is a physics-informed machine learning
(PIML) framework for predictive turbulence modeling. {Here, ``physics-informed'' is to 
emphasize the attempt of accounting for the physical domain knowledge in every stage of 
machine learning.} The problem targeted by the PIML framework can
be formulated as follows: given high-fidelity data (e.g., Reynolds stresses from DNS or resolved
LES) from a set $\{\mathcal{T}_i\}_{i = 1}^{N}$ of $N$ training flows, the framework shall allow for
using standard RANS turbulence models to predict a new flow $\mathcal{P}$ for which data are not
available. The flows $\mathcal{T}_i$ for which high-fidelity simulation data are available are
referred to as \emph{training flows}, and the flow $\mathcal{P}$ to be predicted is referred to as
\emph{test flow}.  The lack of data in test flows is typical in industrial CFD simulations performed
to support design and optimization. Furthermore, we assume that the training flows and the test
flow have similar complexities and are dominated by the same characteristics such as separation or
shock--boundary layer interaction. {This scenario is common in the engineering design process, 
where the test flow is closely related to the training flows. Ultimately, the envisioned machine learning 
framework will be used in scenarios where the training flows consist of a wide range of elementary 
and complex flows with various characteristics and the test flow has a subset or all of them. 
However, the latter scenario is much more challenging and is outside the scope of the current study.
Considering the proposed method is a completely new paradigm, we decide to take small steps by 
starting from the closely related flows and to achieve the overarching goal gradually. }

\subsection{Summary of Proposed Approach}
In the proposed approach we utilize training data to construct functions of the discrepancies
(compared to the DNS data) in the RANS-predicted Reynolds stresses and use these functions to
predict Reynolds stresses in new flows. The procedure is summarized as follows:
 \begin{enumerate}
 \item Perform baseline RANS simulations on both the training flows and the test flow.
 \item Compute the feature vector field $\mathbf{q}(\mathbf{x})$, e.g., pressure gradient and
   streamline curvature, based on the RANS-predicted mean flow fields for all flows.
 \item Compute the discrepancies field $\Delta \tau_\alpha(\mathbf{x})$ in the RANS modeled Reynolds
   stresses for the training flows based on the high-fidelity data.
 \item Construct regression functions $ f_\alpha: \mathbf{q} \mapsto \Delta \tau_\alpha$ for the
   discrepancies based on the training data prepared in Step 3.
 \item Compute the Reynolds stress discrepancies for the test flow by querying the regression
   functions. The Reynolds stresses can subsequently be obtained by correcting the baseline RANS
   predictions with the evaluated discrepancies.
 \end{enumerate}


 In machine learning terminology the discrepancies $\Delta \tau_\alpha$ here are referred to as
 \emph{responses} or \emph{targets}, the feature vector $\mathbf{q}$ as \emph{input}, and the
 mappings $ f_\alpha: \mathbf{q} \mapsto \Delta \tau_\alpha $ as \emph{regression functions}.  A
 regression function $f_\alpha$ maps the input feature vector $\mathbf{q}$ to the response $\Delta
 \tau_\alpha$, and the term ``function'' shall be interpreted in a broad sense here. That is,
 depending on the regression technique used, it can be either deterministic (e.g., for linear
 regression) or random (e.g., Gaussian process)~\cite{friedman2001elements,rasmussen2006gaussian},
 and it may not even have an explicit form. In the case of random forests regression used in this
 work~\cite{breiman2001random}, the mapping does not have an explicit expression but is determined
 based on a number of decision trees.

 In the procedure described above, after the baseline RANS simulations in Step 1, the input feature
 fields are computed in Step 2, the training data are prepared in Step~3, and the regression
 functions are constructed in Step 4. Finally, the regression functions are evaluated to make
 predictions in Step~5. {It is worth noting that in each stage domain knowledge is
 incorporated, e.g., physical reasoning for identification of input features and consideration
 of realizability constraints of Reynolds stress in learning-prediction process.} Each component is 
 discussed in detail below. The choice of features and responses are presented in 
 Sections~\ref{sec:feature} and~\ref{sec:discrepancy}, respectively, and
 the machine learning algorithm chosen to build the regression function is introduced in
 Section~\ref{sec:regression}.

\subsection{Choice of Mean Flow Features as Regression Input}
\label{sec:feature}

As has been pointed out in Section~\ref{sec:intro}, mean flow features are better suited as input of
the regression function than physical coordinates as they allow the constructed functions to predict
flows in different geometries. Ling and Templeton~\cite{ling2015evaluation} proposed a rich set of
twelve features based on clear physical reasoning. The set of features used in the present study
mostly follow their work, except that we excluded the feature ``vortex stretching'' (input 8 in
Table II of ref.~\cite{ling2015evaluation}).  This feature is present only in three-dimensional
flows, but the test cases presented here are two-dimensional flow.  We excluded two additional
features related to linear and nonlinear eddy viscosities (features 6 and 12 in
ref.~\cite{ling2015evaluation}). These quantities were specifically chosen for evaluating
qualitative confidence indicators of RANS predictions and, in our opinion, are not suitable input
for regression functions of Reynolds discrepancies.  Finally, experiences in the turbulence modeling
communities suggest that mean streamline curvature has important influences on the predictive
performance of RANS models~\cite{spalart2000strategies}. Therefore, curvature is included as an
additional feature.  The complete list of the mean flow features chosen as regression inputs in this
work is summarized in Table~\ref{tab:feature}.

In choosing the mean flow features as regression inputs, we have observed a few principles in
general. First, the input and thus the obtained regression functions should be
Galilean-invariant. Quantities that satisfy this requirement include all scalars and the invariants
(e.g., norms) of vectors and tensors. An interesting example is the pressure gradient along
streamline (see feature~$q_4$ in Table~\ref{tab:feature}).  While neither velocity $U_k$ nor pressure
gradient $dP/d x_k$ (both being vectors) is Galilean-invariant by itself and thus is not a suitable
input, their inner product $U_k \frac{dP}{d x_k}$ is.  Second, since the truth of the mean flow
fields in the test flows are not available, an input should solely utilize information of the mean
flow field produced by the RANS simulations~\cite{ling2015evaluation}. Therefore, all of the adopted
features are based on RANS-predicted pressure $P$, velocity $\mathbf{U}$, turbulent kinetic
energy~$k$, and distance $d$ to the nearest wall. Finally, to facilitate implementation and avoid
ambiguity, only local quantities (i.e., cell- or point-based quantities in CFD solvers) of the flow
field are used in the formulation of features, with the distance $d$ to nearest wall being a notable
exception.  This principle is similar to that in choosing variables for developing turbulence
models~\cite{spalart2000strategies}.

The interpretation of most feature variables are evident from the brief descriptions given in the
table, but a few need further discussions. First, feature $q_1$ (Q-criterion) is based on the
positive second invariant $Q$ of the mean velocity gradient $\nabla \mathbf{U}$, which represents
excess rotation rate relative to strain rate~\cite{hunt1988eddies}. For incompressible flows, it can
be computed as $Q=\frac{1}{2}(\|\boldsymbol{\Omega}\|^2 - \|\mathbf{S}\|^2)$, where
$\boldsymbol{\Omega}$ and $\mathbf{S}$ are rotation rate and strain rate tensors, respectively;
$\|\boldsymbol{\Omega} \| = \sqrt{\operatorname{tr}(\boldsymbol{\Omega}\boldsymbol{\Omega}^T)}$ and
$\|\mathbf{S}\| = \sqrt{\operatorname{tr}(\mathbf{S} \mathbf{S}^T)}$, with superscript $T$
indicating tensor transpose and tr indicating trace. The Q-criterion is widely used in CFD
simulations as a post-processing tool to identify vortex structures for the visualization of flow
structures~\cite{chakraborty2005relationships}. Second, the wall distance based Reynolds number
$Re_d = \sqrt{k} d/ \nu$ in $q_3$ is an indicator to distinguish boundary layers from shear
flows. This is an important feature because RANS models behave very differently in the two types of
flows. This quantity is frequently used in wall functions for turbulence models.  Third, feature
$q_7$ defines the deviation from orthogonality between the velocity and its
gradient~\cite{gorle2012rans}, which indicates the deviation of the flow from parallel shear flows
(e.g., plane channel flows). Since most RANS models are calibrated to yield good performance on
parallel shear flows, this deviation is usually correlated well with large discrepancies in RANS
predictions. However, since it only accounts for misalignment angle and not the velocity magnitude,
in regions with near-zero velocities this quantity becomes the angle formed by two zero-length
vectors and is thus mostly noise. Finally, we remark that most of the features in
Table~\ref{tab:feature} including the Q-criterion and wall-distance based Reynolds number are
familiar to CFD practitioners.

Most of the features presented in Table~\ref{tab:feature} are formulated as ratios of two quantities
of the same dimension, either explicitly ($q_1$, $q_5$, $q_6$, $q_8$, $q_9$) or implicitly ($q_2$
and $q_3$). Hence, they are non-dimensional by construction.  Features $q_4$ and $q_7$ involve inner
product of vectors or tensors, and thus they are normalized by the magnitude of the constituent
vectors or tensors.  Finally, feature $q_{10}$ (streamline curvature) is normalized by $1/L_c$,
where $L_c$ is the characteristics length scale of the mean flow, chosen to be the hill height $H$
(see Fig.~\ref{fig:flowGeo}) in the numerical examples.

We followed the procedure of Ling and Templeton~\cite{ling2015evaluation} to normalize the features.
Except for feature $q_3$, each element $q_\beta$ in the input vector $\mathbf{q}$ is normalized as:
\begin{equation}
  \label{eq:normlize-feature}
  q_\beta =  \frac{\hat{q}_\beta}{|\hat{q}_\beta| + |q_\beta^*|} \quad  \textrm{where} \quad
  \beta = 1,  2, 4, \cdots, 10, 
\end{equation}
where the summation on repeated indices is not implied, $\hat{q}_\beta$ are raw values of the features, and
${q_\beta^*}$ are the corresponding normalization factors. This normalization scheme limits the
numerical range of the inputs within $[-1, 1]$ and thus facilitates the regression.  The normalization
is not needed for feature $q_3$ (wall distance based Reynolds number) since it is already in a
non-dimensional form and in a limited range $[0, 2]$.

It can be seen that the choices of features and normalization factors heavily rely on physical
understanding of the problem (turbulence modeling). That is, in the present data-driven modeling
framework, the data are utilized only \emph{after} physical reasoning from the modeler has been
applied. This task can be a burden in certain applications. {It is worth noting that the recent work
of Ling et al.~\cite{ling2016jcp} aimed to relieve the modeler from such burdens by using a basis
of invariants of tensors relevant in the specific application (e.g., strain rate $\mathbf{S}$ in 
turbulence modeling). Their work has the potential to systematically construct the input
features based on raw physical variables and thus makes data-driven modeling even ``smarter''.}

\begin{table}[htbp] 
  \centering
  \caption{
    Non-dimensional flow features used as input in the regression.  The normalized feature $q_\beta$ is
    obtained by normalizing the corresponding raw features value $\hat{q}_\beta$ with normalization
    factor $q^*_\beta$ according to $q_\beta = \hat{q}_\beta / (|\hat{q}_\beta| + |q^*_\beta|)$
    except for $\beta = 3$. Repeated indices imply summation  for indices
    $i$, $j$, $k$, and $l$ but not for $\beta$. Notations are as follows: $U_i$ is mean velocity, $k$
    is turbulent kinetic energy (TKE), $u'_i$ is fluctuation velocity, $\rho$ is fluid density,
    $\varepsilon$ is the turbulence dissipation rate, $\mathbf{S}$ is the strain rate tensor,
    $\boldsymbol{\Omega}$ is the rotation rate tensor, $\nu$ is fluid viscosity, $d$ is distance to
    wall, $\boldsymbol{\Gamma}$ is unit tangential velocity vector,  $D$ denotes material derivative,
    and $L_c$ is the characteristic length scale of the mean flow. $\| \cdot \|$ and 
    $|\cdot|$ indicate matrix and vector norms, respectively.  }
\label{tab:feature}
\begin{tabular}{P{2.0cm} | P{5.0cm}  P{3.0cm}  P{4.0cm} }	
  \hline
  feature ($q_\beta$)  & description & raw feature ($\hat{q}_\beta$) &
  normalization factor ($q^*_\beta$)  \\ 
  \hline
  $q_1$  & ratio of excess rotation rate to strain rate (Q-criterion) &
  $\frac{1}{2}(\|\boldsymbol{\Omega}\|^2 - \|\mathbf{S}\|^2)$ & 
  $\|\mathbf{S}\|^2$\\  
  \hline
  $q_2$  &   turbulence intensity & $k$ &
  $\frac{1}{2}U_iU_i$\\ 
  \hline
  $q_3$  & wall-distance based Reynolds number &
  $\min{\left(\dfrac{\sqrt{k}d}{50\nu}, 2\right)}$ & not applicable$^{(a)}$\\
  \hline
  $q_4$  & pressure gradient along streamline & $U_k\dfrac{\partial P}{\partial
    x_k}$ & $\sqrt{\dfrac{\partial P}{\partial x_j} \dfrac{\partial P}{\partial
      x_j}U_iU_i}$ \\ 
  \hline
  $q_5$  & ratio of turbulent time scale  to mean strain time scale &{ $\dfrac{k}{\varepsilon}$ }
  & { $\dfrac{1}{\|\mathbf{S}\|} $ } \\ 
  \hline
  $q_6$  & ratio of pressure normal stresses to shear stresses &
  $\sqrt{\dfrac{\partial P}{\partial x_i}\dfrac{\partial P}{\partial x_i}}$ &
  $\dfrac{1}{2} \rho\dfrac{\partial U_k^2}{\partial x_k}$\\ 
  \hline
  $q_7$  & non-orthogonality between velocity and its gradient~\cite{gorle2012rans} & $\left| U_i U_j\dfrac{\partial
      U_i}{\partial x_j} \right|$ &{$\sqrt{U_l U_l \, U_i \dfrac{\partial U_i}{\partial
        x_j}U_k \dfrac{\partial U_k}{\partial x_j}}$ } \\ 
  \hline
  $q_8$  & ratio of convection to production of TKE & $U_i\dfrac{dk}{dx_i}$ &
  $|\overline{u'_j u'_k} S_{j k}|$\\ 
  \hline
  $q_9$  & ratio of total to normal Reynolds stresses &
  $\|\overline{u'_iu'_j}\|$ & $k$\\ 
  \hline
  $q_{10}$  & streamline curvature & {$\left|\frac{D
        \boldsymbol{\Gamma}}{ D s}\right|$ where $\boldsymbol{\Gamma} \equiv
    \mathbf{U}/|\mathbf{U}|$, $D s = |\mathbf{U}| Dt$ } & { $\dfrac{1}{L_c}$ }\\
  \hline						 								
\end{tabular}
\flushleft
{\small
  Note: (a)  Normalization is not necessary as the Reynolds number is non-dimensional.}
\end{table}



\subsection{Representation of Reynolds Stress Discrepancies as Responses} 
\label{sec:discrepancy}

In Section~\ref{sec:motivation} the Reynolds stress discrepancies $\Delta \tau$ have been identified
as the responses for the regression functions.  The response quantities should also be based on
Galilean invariant quantities due to the same consideration as in the feature choice. As such,
individual components of the Reynolds stresses or the discrepancies based thereon are not suitable,
but those based on their eigenvalues or invariants are preferred. In turbulence modeling, the Lumley
triangle has been widely used for the analysis of turbulence states related to
realizability~\cite{lumley1977return}. It is formulated based on the second and third invariants
($II$ and $III$) of the anisotropy tensor. Recently, Banerjee et al.~\cite{banerjee2007presentation}
proposed an improved formulation in which the eigenvalues of the anisotropy tensor are mapped to
Barycentric coordinates as opposed to the variants $II$ and $III$ as in the Lumley triangle. {An
important advantage of their formulation is that the mapping to Barycentric coordinates is linear,
which is in contrast to the nonlinear mapping to invariants $II$ and $III$. Therefore, Barycentric
coordinate provides a non-distorted visual representation of anisotropy and is easier to impose
realizability constraints.}  The formulation of discrepancy starts with the eigen-decomposition of 
the Reynolds stress anisotropy tensor $\mathbf{A}$:
\begin{equation}
  \label{eq:tau-decomp}
  \boldsymbol{\tau} = 2 k \left( \frac{1}{3} \mathbf{I} +  \mathbf{A} \right)
  = 2 k \left( \frac{1}{3} \mathbf{I} + \mathbf{V} \Lambda \mathbf{V}^T \right)
\end{equation} 	
where $k$ is the turbulent kinetic energy, which indicates the magnitude of $\boldsymbol{\tau}$; $\mathbf{I}$
is the second order identity tensor; $\mathbf{V} = [\mathbf{v}_1, \mathbf{v}_2, \mathbf{v}_3]$ and
$\Lambda = \textrm{diag}[\lambda_1, \lambda_2, \lambda_3]$ with $\lambda_1+\lambda_2+\lambda_3=0$
are the orthonormal eigenvectors and eigenvalues of $\mathbf{A}$, respectively, indicating its shape
and orientation.

\begin{figure}[!htbp]
\centering
\includegraphics[width=0.5\textwidth]{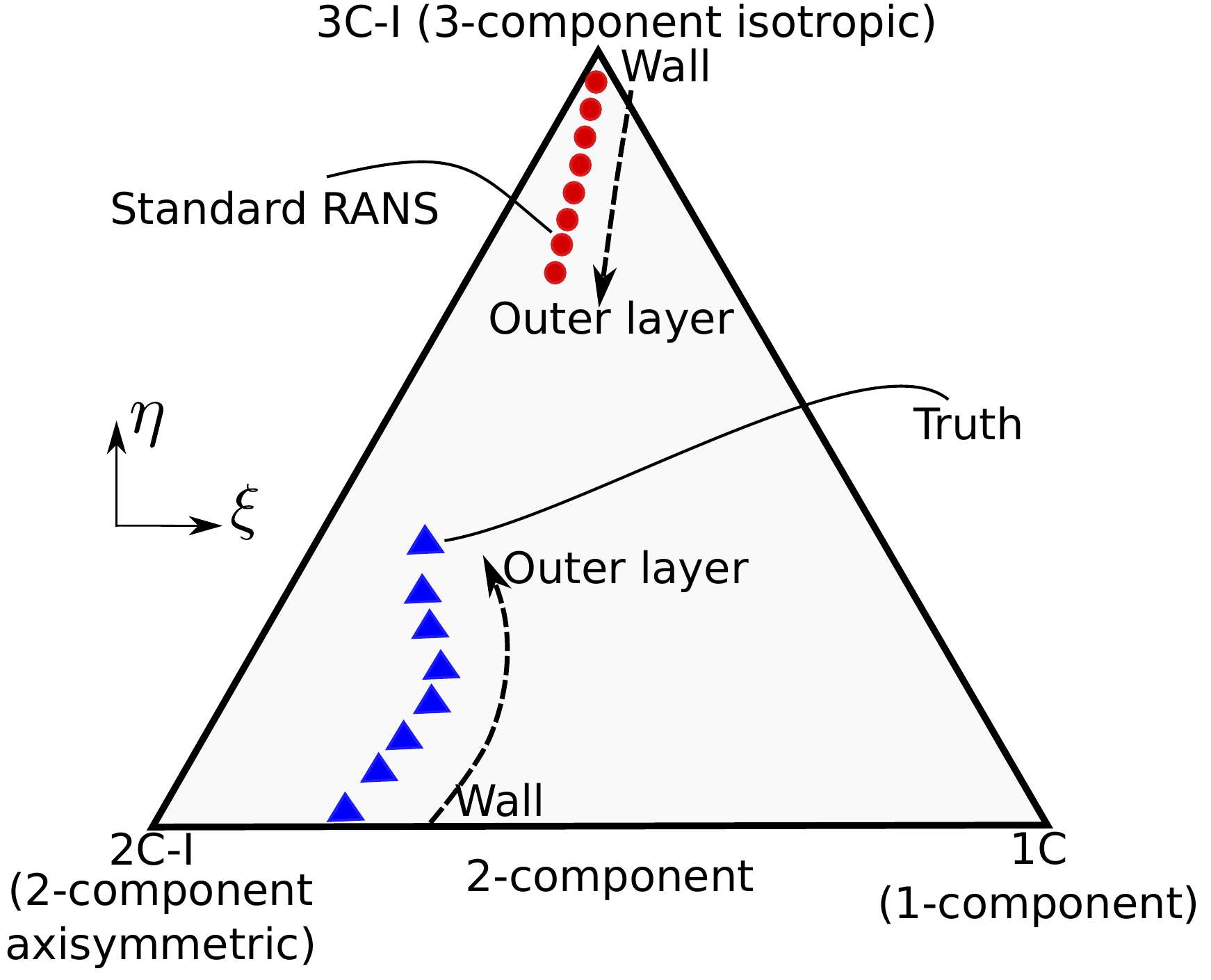}
\caption{ The Barycentric triangle that encloses all physically realizable states of Reynolds
  stress~\cite{banerjee2007presentation,emory2014componentality}. The position within the Barycentric
  triangle represents the anisotropy state of the Reynolds stress.  Typical mapped locations of near
  wall turbulence states are indicated with prediction from standard RANS models near the isotropic
  state (vertex 3C-I) and the turth near the bottom edge (2C-I). The typical RANS-predicted trends of
  spatial variation from the wall to shear flow and the corresponding truth are indicated with
  arrows.}
\label{fig:bary}
\end{figure}

In the Barycentric triangle, the eigenvalues $\lambda_1$, $\lambda_2$, and $\lambda_3$ are mapped to
the Barycentric coordinates $(C_1,C_2, C_3)$ as follows:
\begin{subequations}
	\label{eq:lambda2c}
        \begin{align}
          C_1 & = \lambda_1 - \lambda_2 \\
          C_2 & = 2(\lambda_2 - \lambda_3) \\
          C_3 & = 3\lambda_3 + 1
        \end{align}  
\end{subequations}
with $C_1 + C_2 + C_3 = 1$.  As shown in Fig.~\ref{fig:bary}, the Barycentric coordinates of a point
indicate the portion of areas of three sub-triangles formed by the point and with edges of
Barycentric triangle.  For example, a point located on the top vertex corresponds to $C_3 = 1$ while
a point located on the bottom edge has $C_3$ = 0.  Similar to the Lumley triangle, all realizable
turbulences are enclosed in the Barycentric triangle (or on its edges) and have positive
Barycentric coordinates $C_1$, $C_2$, and $C_3$.  The Barycentric triangle has been used by Emory et
al.~\cite{emory2011modeling} as a mechanism to impose realizability of Reynolds stresses in
estimating uncertainties in RANS simulations.

Placing the triangle in a Cartesian coordinate system $\boldsymbol{\xi} \equiv (\xi, \eta)$, the
location of any point within the triangle is a convex combination of those of the three vertices,
i.e.,
\begin{equation}
\boldsymbol{\xi} = 	\boldsymbol{\xi}_{1c}C_1 + \boldsymbol{\xi}_{2c}C_2 +
\boldsymbol{\xi}_{3c}C_3 
\end{equation}	
where $\boldsymbol{\xi}_{1c}$, $\boldsymbol{\xi}_{2c}$, and $\boldsymbol{\xi}_{3c}$ denote
coordinates of the three vertices of the triangle. An advantage of representing the anisotropy of
Reynolds stress in the Barycentric coordinates is that it has a clear physical interpretation, i.e.,
the dimensionality of the turbulence state~\cite{emory2014componentality}.  Typically,
the standard-RANS-predicted Reynolds stress at a near wall location is located close to the
isotropic, three-component state (vertex 3C-I) in the Barycentric triangle, while the truth is near
the two-component limiting states (bottom edge).  Moreover, the spatial variations from the near-wall
region to the shear flow region are indicated as arrows in Fig.~\ref{fig:bary}.  It is clear that
the trend of spatial variation predicted by a standard RANS model is opposite to that of the truth.

The three mutually orthogonal, unit-length eigenvectors $\mathbf{v}_1$, $\mathbf{v}_2$, and
$\mathbf{v}_3$ indicate the orientation of the anisotropy tensor. They can be considered a rigid
body and thus its orientation has three degrees of freedom, although they have nine elements in
total.  We use the Euler angle with the $z$--$x'$--$z''$ convention to parameterize the orientation
following the convention in rigid body dynamics~\cite{goldstein1980euler}. That is, if a local
coordinate system $x$--$y$--$z$ spanned by the three eigenvectors of $\mathbf{V}$ was initially
aligned with the global coordinate system ($X$--$Y$--$Z$), the current configuration could be
obtained by the following three consecutive intrinsic rotations about the axes of the local
coordinate system: (1) a rotation about the $z$ axis by angle $\varphi_1$, (2) a rotation about the
$x$ axis by $\varphi_2$, and (3) another rotation about its $z$ axis by $\varphi_3$.  The local
coordinate axes usually change orientations after each rotation. 

In summary, the Reynolds stress tensor is projected to six physically interpretable, Galilean
invariant quantities representing the magnitude ($k$), shape ($\xi$, $\eta$), and orientation
($\varphi_1$, $\varphi_2$, $\varphi_3$). They are collectively denoted as $\tau_\alpha$.  The truths
of these quantities can be written as baseline RANS predictions corrected by the corresponding
discrepancy terms, i.e.,
\begin{subequations}
	\label{eq:delta}
	\begin{alignat}{2}
	\log k & = &\ \log \tilde{k}^{rans}  & + \Delta \log k\\
	\xi & = &\ \tilde{\xi}^{rans} & + \Delta \xi  \\
	\eta & = &\ \tilde{\eta}^{rans} & + \Delta \eta\\
	\varphi_{i} & = &\ \tilde{\varphi_i}^{rans} & + \Delta \varphi_i, \; \textrm{for} \; \  i = 1, 2, 3.  
	\label{eq:kdelta}    
	\end{alignat}
\end{subequations}
The discrepancies ($\Delta \log k$, $\Delta \xi$,
$\Delta \eta$, $\Delta \varphi_1$, $\Delta \varphi_2$, $\Delta \varphi_3$, denoted as $\Delta
\tau_\alpha$ with $\alpha = 1, 2, \cdots, 6$) in the six projections of the Reynolds stress tensor
are responses of the regression functions.  We utilize data consisting of pairs of $(\mathbf{q}, \Delta
\tau_\alpha)$ from training flow(s) to construct the functions $f_\alpha: \mathbf{q} \mapsto \Delta
\tau_\alpha$.  It is assumed that the discrepancies in six quantities $\Delta \tau_\alpha$ are
independent, and thus separate functions are built for each of them. This simplification is
along the same lines as that made in previous works~\cite{tracey2013application}.

\subsection{Random Forests for Building Regression Functions}
\label{sec:regression}
With the input (mean flow features $\mathbf{q}$) and responses (Reynolds stress discrepancies
$\Delta\tau_\alpha$) identified above, a method is needed to construct regression functions from
training data and to make predictions based on these functions. Supervised machine learning consists
of a wide variety of such methods including K-nearest neighbors~\cite{altman1992introduction}, linear regression and its variants
(e.g., Lasso)~\cite{james2013introduction}, Gaussian processes~\cite{rasmussen2006gaussian},
tree-based methods (decision trees, random forests, bagging)~\cite{breiman2001random}, neural
networks~\cite{anderson1995introduction}, and support vector machine~\cite{cristianini2000introduction}, among others. A major consideration in choosing the regression
method is the high dimensionality of the feature space, which is typically ten or higher in our
application. The curse of dimensionality makes such methods as K-nearest neighbors, linear
regression, and Gaussian processes not suitable. Secondary considerations, which we believe are also
important for turbulence modeling applications, are the capability to provide predictions with
quantified uncertainties as well as physical insights (e.g., on the importance of each of the features and
their interactions).  After evaluating a number of existing machine learning techniques in light of
these criteria, we identified \emph{random forests}~\cite{breiman2001random} as the optimal approach
for our purposes, which is an ensemble learning technique based on decision trees.

In simple decision tree learning, a tree-like model is built to predict the response variable by
learning simple if-then-else decision rules from the training data. Decision trees have the
advantage of being easy to interpret (e.g., via visualization) and implement. They are also
computationally cheap. However, they tend to overfit the data and lack robustness. That is, a small
change in the training data can result in large changes in the built model and its predictions.
Random forests learning is an ensemble learning technique proposed by Ho~\cite{ho1998random} and
Briemann~\cite{breiman1984classification,breiman2001random} that overcomes these shortcomings of
simple decision trees. Since these techniques are generally not familiar to readers in the fluid
dynamics community, here we use an illustrative example in the context of turbulence modeling to
explain the algorithm.

\begin{figure}
	\centering 
	\subfloat[feature space
	stratification]{\includegraphics[width=0.47\textwidth]{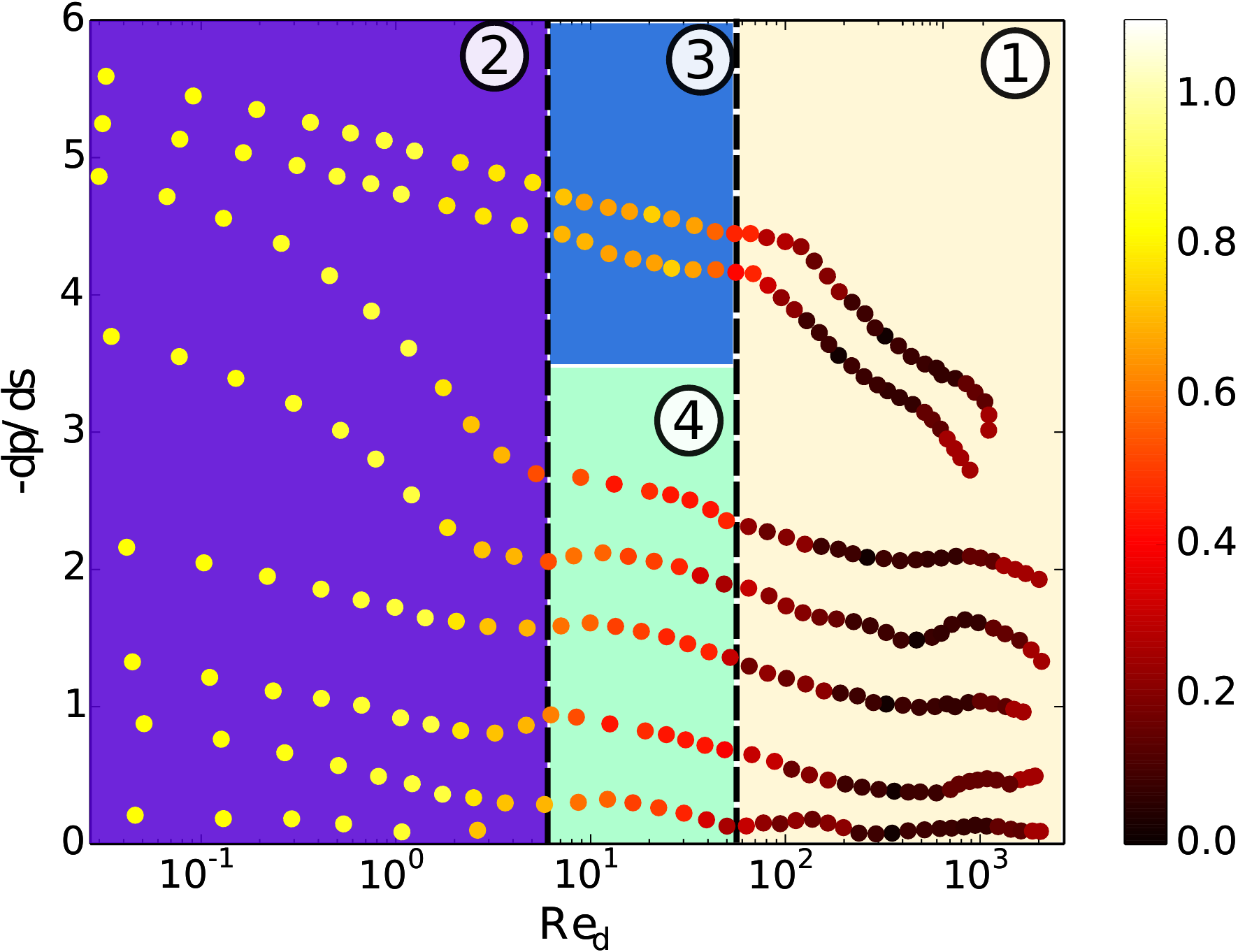}}
	\hspace{1em} %
	\subfloat[regression tree]{\includegraphics[width=0.45\textwidth]{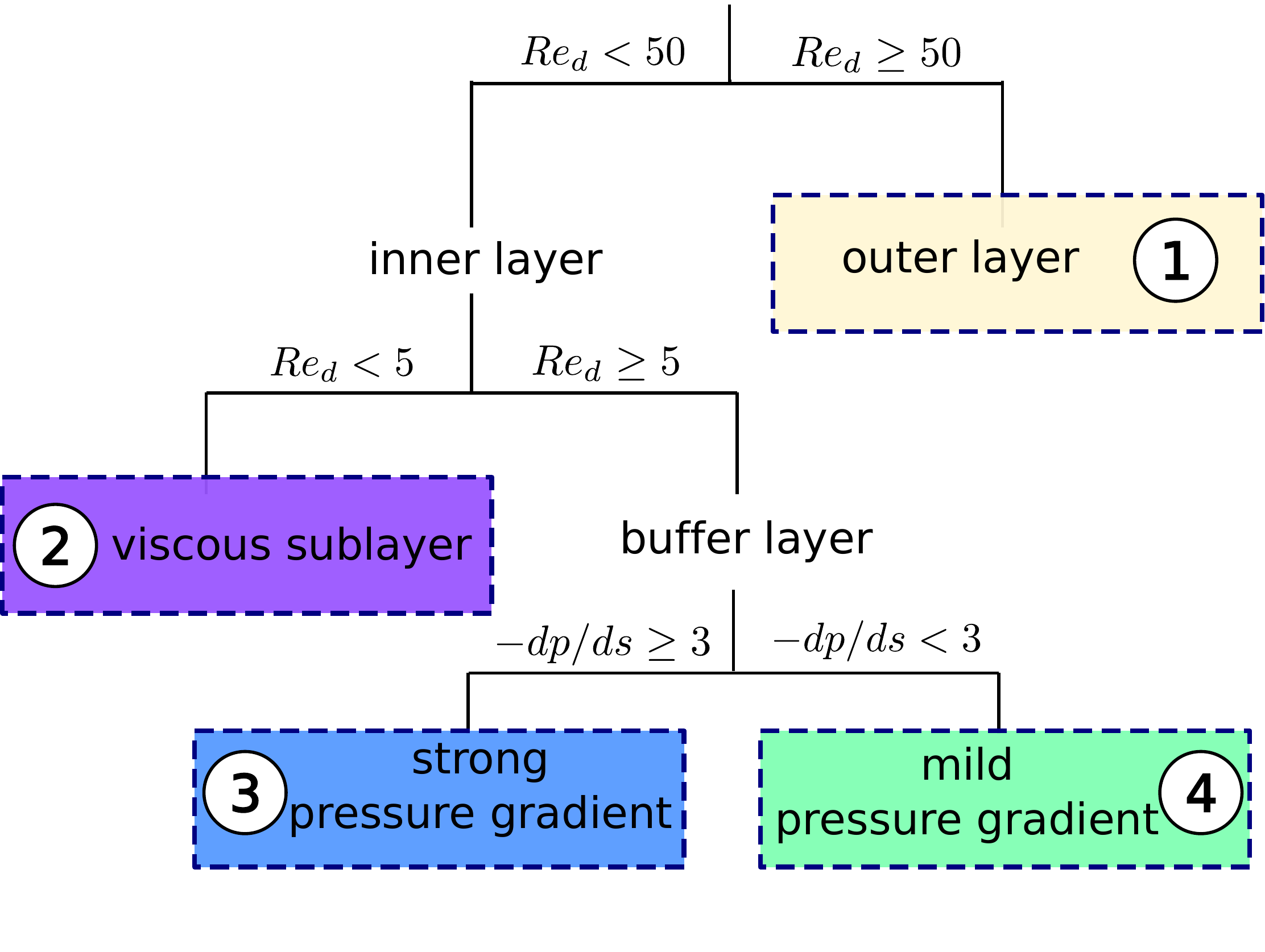}} %
	\caption{Schematic of a simple regression tree in a two-dimensional feature space (pressure
		gradient along streamline $dp/ds$ and wall-distance based Reynolds number $Re_d$), showing (a)
		the stratification of feature space and (b) the corresponding regression tree built from the
		training data. The response is the discrepancy $\Delta \eta$ in the Barycentric triangle of the
		RANS-predicted Reynolds stress. When predicting the discrepancy for a given feature vector
		$\tilde{\mathbf{q}}$, the tree model in (b) is traversed to identify the leaf, and the mean of
		the training data is taken as the prediction $\Delta \eta (\tilde{\mathbf{q}})$.}
	\label{fig:tree-schematic}
\end{figure}

A simple decision tree model is illustrated in Fig.~\ref{fig:tree-schematic}. For clarity we
consider an input with only two features: pressure gradient $dp/ds$ (normalized and projected to the
streamline tangential) and wall distance based Reynolds number $Re_d = C_\mu^{1/4} d \sqrt{k}/\nu$,
as defined in Table~\ref{tab:feature}. It can be also interpreted as wall distance in viscous unit.
The response is the discrepancy $\Delta \eta$ of the vertical coordinate in the Barycentric triangle
of the RANS-predicted Reynolds stress (see Fig.~\ref{fig:bary}).  During the training process, the
feature space is \emph{successively} divided into a number of boxes (leafs) based on the training
data (shown as points in the $dp/ds$--$Re_d$ plane in Fig.~\ref{fig:tree-schematic}a).  In the
simplest decision tree model used for regression, the feature space is stratified with the objective
of minimizing the total in-leaf variances of the responses at each step, a strategy that is referred
to as greedy algorithm.  After the stratification, a constant prediction model is built on each
leaf.  When predicting the response $\Delta \eta$ for a given feature vector $\mathbf{q}$, the
constructed tree model in Fig.~\ref{fig:tree-schematic}b is traversed to identify the leaf where
$\mathbf{q}$ is located, and the mean response on the leaf is taken as the prediction $\Delta \eta
(\mathbf{q})$.

The tree model has a clear physical interpretation in the context of turbulence modeling. For
example, it is well known that a standard isotropic eddy viscosity model has the largest discrepancy
when predicting anisotropy in the viscous sublayer ($Re_d \le 5$). This is because the truth is
located on the bottom, corresponding to a combination of one- or two-component turbulence, while a
typical isotropic eddy viscosity model would predict an isotropic state located on the top vertex
(see Fig.~\ref{fig:bary}). In contrast, far away from the wall within the outer layer ($Re_d > 50$),
the RANS-predicted anisotropy is rather satisfactory. Therefore, the first two branches divide the
space to three regions based on the feature $Re_d$: outer layer (region 1) , viscous sublayer
(region 2), and buffer layer (regions 3 and 4). In the buffer layer the pressure gradient plays a
more important role than in the outer and viscous layers. Larger pressure gradients correspond to
larger discrepancies, which can be explained by the fact that favorable pressure gradients (negative
$dp/ds$ values) tend to thicken the viscous sublayer~\cite{jones1972prediction}, which leads to
larger discrepancies in $\eta$. Therefore, a further division splits the buffer layer states to two
regions in the feature space, i.e., those with strong (region 3) and mild (region 4) pressure
gradients.

A simple regression tree model described above tends to overfit for high dimensional input space,
i.e., yielding models that explain the training data very well but predict poorly for unseen
data. In general the decision trees do not have the same level of predictive accuracy as other modern
regression methods. However, by aggregating a large number of trees (ideally with minimum
correlation), the predictive performance can be significantly improved and, the overfitting can be largely
avoided.  In random forests an ensemble of trees is built with bootstrap samples (i.e., sampling
with replacement) drawn from the training data~\cite{friedman2001elements}.  Moreover, when
building each tree, it utilizes only a subset of $M \le N_q$ randomly chosen features among the
$N_q$ features, which reduces the correlation among the trees in the ensemble and thus decreases the
bias of the ensemble prediction.

Random forest regression is a modern machine learning method with predictive performance comparable
to other state-of-the-art techniques~\cite{james2013introduction}.  In decision tree models the
maximum depth of trees must be limited (e.g., by pruning the branches far from the root) to ensure a
sufficient number of training point (e.g., 5) on each node. In contrast, in random forests, one can
build each tree to its maximum depth by successive splitting the nodes until only one training data
point remains on each leaf.  While each individual tree built in this manner may suffer from
overfitting and has large prediction variances, the use of ensemble largely avoids both
problems. Moreover, random forest regression is simple to use with only two free parameters, i.e.,
the number $N_{rf}$ of trees in the ensemble and the number $M$ of selected features. In this work
we used an ensemble of $N_{rf} = 100$ trees and the a subset of features (i.e., $M = 6$) to
build each tree. As a standard practice in statistical modeling, we performed cross-validations to
optimize these parameters and performed sensitivity analysis to ensure that the predictions are not
sensitive to the parameter choices.

\section{Numerical Results}
\label{sec:result}
Almost all industrial flows involve some characteristics (e.g., strong pressure gradient, 
streamline curvature, and separation) that break the equilibrium assumption of RANS model. 
Therefore, we have these challenges in mind when developing the data-driven approach.
{In this study, we focus on the cases where training and test flows have similar 
	characteristics. Specifically, we evaluate the proposed method on two classes of 
	flows: (1) fully developed turbulent flows in a square duct at various Reynolds numbers 
	and (2) flows with massive separations. The flow in a square duct at Reynolds 
	number $Re = 3500$ and the flow in a channel with periodic hills at Reynolds 
	number $Re = 10595$ are chosen as the prediction (test) flows for the respective flow classes. 
	The square duct flow has an in-plane secondary flow pattern induced by the normal 
	stress imbalance, while the periodic-hill flow features a recirculation bubble, non-parallel 
	shear layer and mean flow curvature. All these characteristics are known to pose challenges 
	for RANS based turbulence models, and thus large model-form discrepancies exist in the 
	RANS-modeled Reynolds stresses. In the two test flows, the relative importance of Reynolds stress
	projections to the mean flow prediction are different. The Reynolds stress anisotropy 
	plays an important role in obtaining the accurate secondary mean motion in the 
	duct flow~\cite{bradshaw1987turbulent}. In contrast, the anisotropy is less important to 
	predict the mean flow in the periodic-hill case, where the turbulent shear stress component 
	is more essential to obtain an accurate mean velocity field~\cite{billard2011}. Therefore, we use these two types of 
	flows to highlight the improvements in the different Reynolds stress components that are 
	important for the predictions of QoIs in the respective flow classes. In both cases, all RANS simulations are performed in an open-source CFD platform, OpenFOAM, 
	using a built-in incompressible flow solver \texttt{simpleFoam}~\cite{weller1998tensorial}.
	Mesh convergence studies have been performed.}
\subsection{Turbulent Flows In a Square Duct}
\label{sec:result:duct}
\subsubsection{Case Setup}
{The fully developed turbulent flow in a square duct is a challenging case for RANS-based
	turbulence models, since the secondary mean motion cannot be captured by linear eddy
	viscosity models (e.g., $k$--$\varepsilon$, $k$--$\omega$), and even the Reynolds stress
	transport models (RSTM) cannot predict it well~\cite{billard2011}. In this test, we aim to improve the 
	RANS-modeled Reynolds stresses of the duct flow at Reynold number $Re = 3500$ by using 
	the proposed PIML approach. The training data are obtained from DNS simulations~\cite{pinelli2010reynolds} of the duct flows in the same geometry but at lower Reynolds numbers $Re = 2200, 2600$ and $2900$.
	The DNS data of the prediction flow ($Re = 3500$) are reserved for comparison and are not 
	used for training.}
\begin{figure}[htbp]
	\centering
	\includegraphics[width=0.8\textwidth]{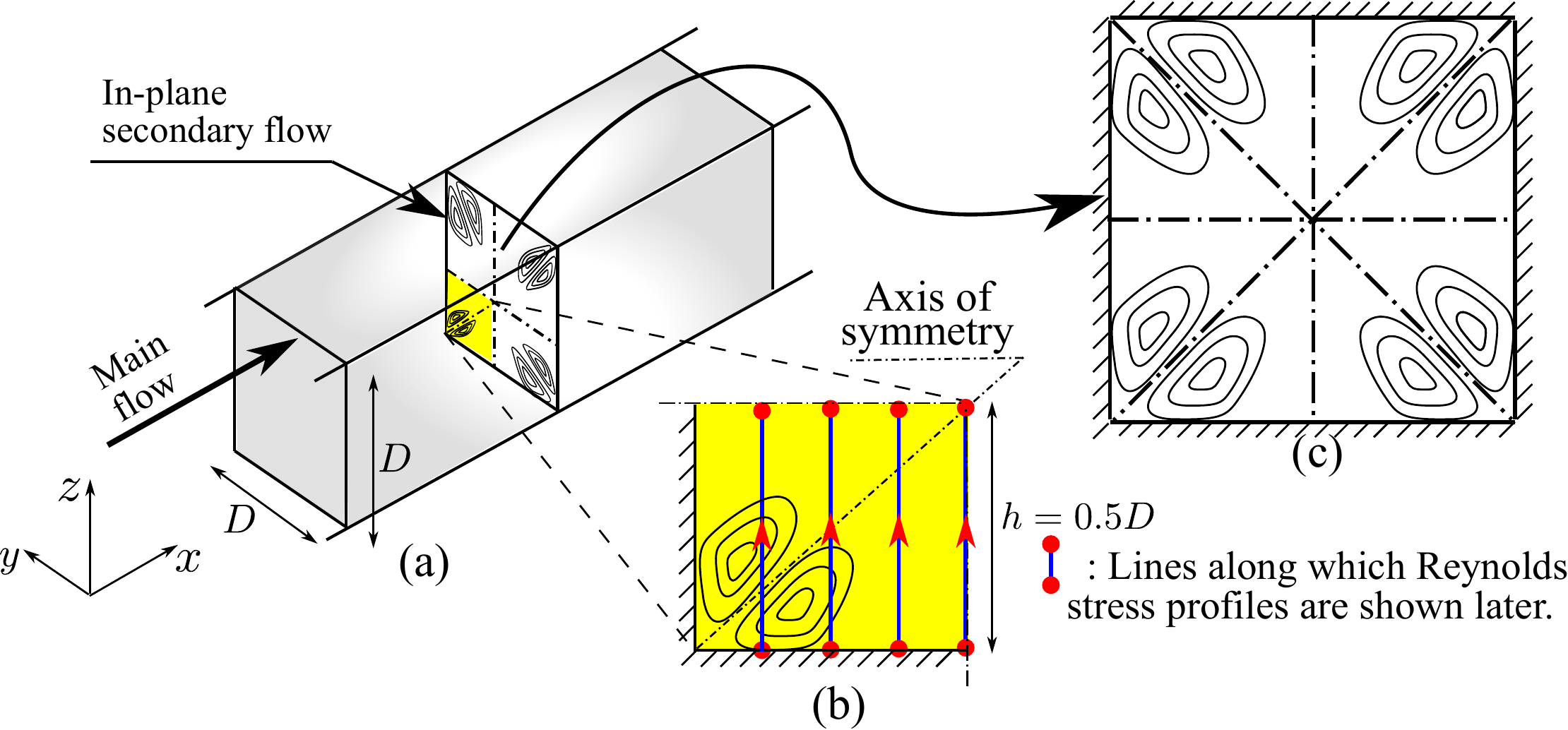}
	\caption{{Domain shape for the flow in a square duct. The $x$ coordinate represents the streamwise
		direction. Secondary flows induced by Reynolds stress imbalance exist in the $y$--$z$
		plane. Panel (b) shows that the computational domain covers a quarter of the cross-section of the
		physical domain. This is due to the symmetry of the mean flow in both $y$ and $z$ directions as
		shown in panel (c).}}
	\label{fig:domain_duct}
\end{figure}
{The geometry of this flow case is shown in~Fig.~\ref{fig:domain_duct}. The Reynolds number is
	based on the edge length D of the square duct and the bulk velocity $U_b$. All lengths presented below are
	normalized by $D/2$.} 
	
{The baseline RANS simulations are performed for all training and test flows. The purpose is 
	twofold: to obtain the mean flow feature fields $\mathbf{q}(\mathbf{x})$ as inputs and to obtain 
	the discrepancies of Reynolds stress by comparing with the DNS data. To enable the comparison, 
	the high-fidelity data are interpolated onto the mesh of RANS simulation. The Launder-Gibson 
	RSTM~\cite{gibson1978ground} is adopted to perform the baseline simulations, since all the 
	linear eddy viscosity models are not able to capture the mean flow features of the secondary motions. 
	The $y^+$ of the first cell center is kept less than 1 and thus no wall model is applied. As indicated 
	in Fig.~\ref{fig:domain_duct}, only one quadrant of the physical domain is simulated owing to the 
	symmetry of the mean flow with respect to the centerlines along $y$- and $z$-axes. No-slip boundary 
	condition is applied on the walls, and symmetry boundary condition is applied on
	the symmetry planes.}

\subsubsection{Prediction Results}
\begin{figure}[htbp]
	\centering 
	\subfloat[$y/H = 0.25$]{\includegraphics[width=0.45\textwidth]{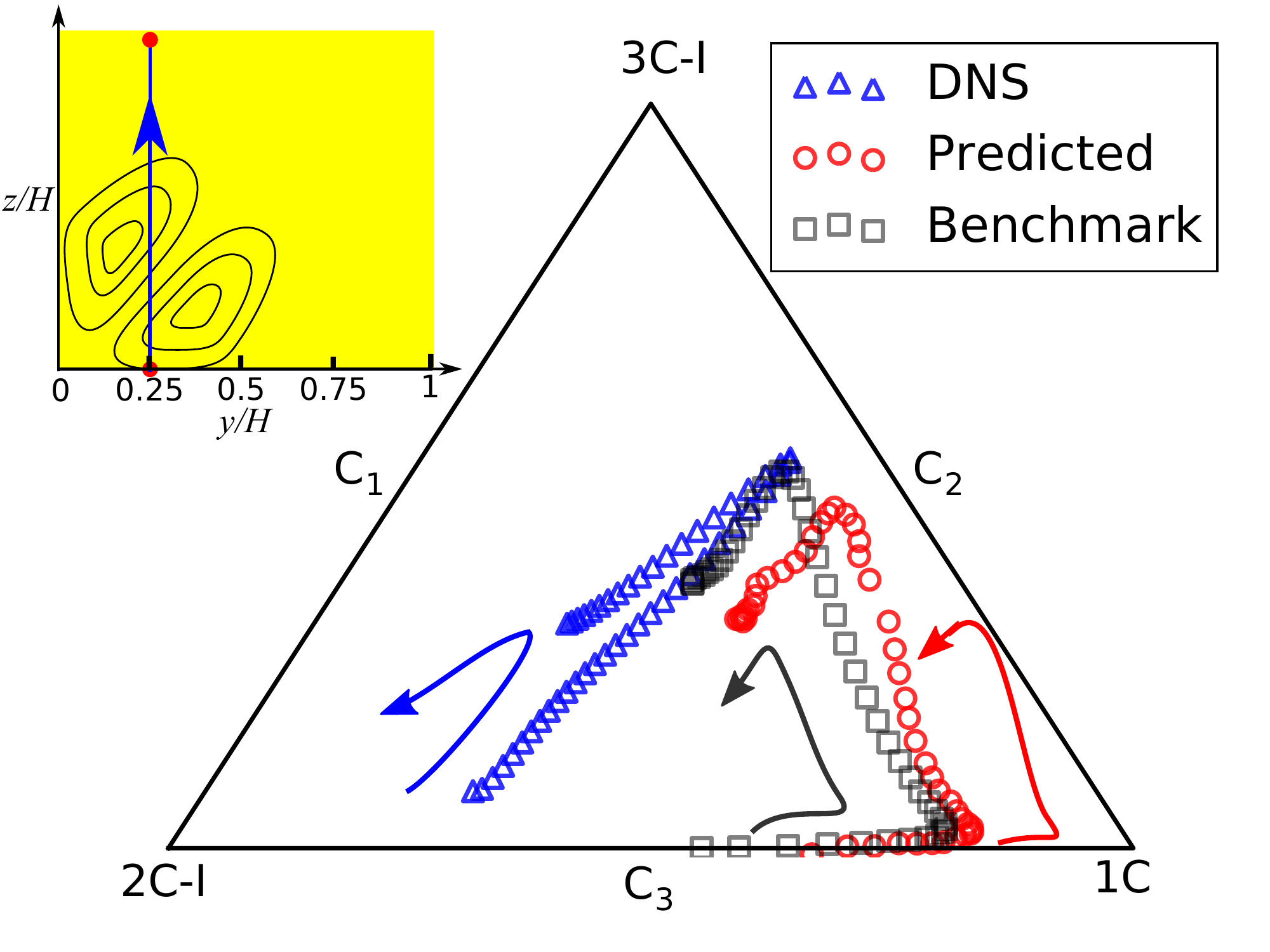}} 
	\subfloat[$y/H = 0.75$]{\includegraphics[width=0.45\textwidth]{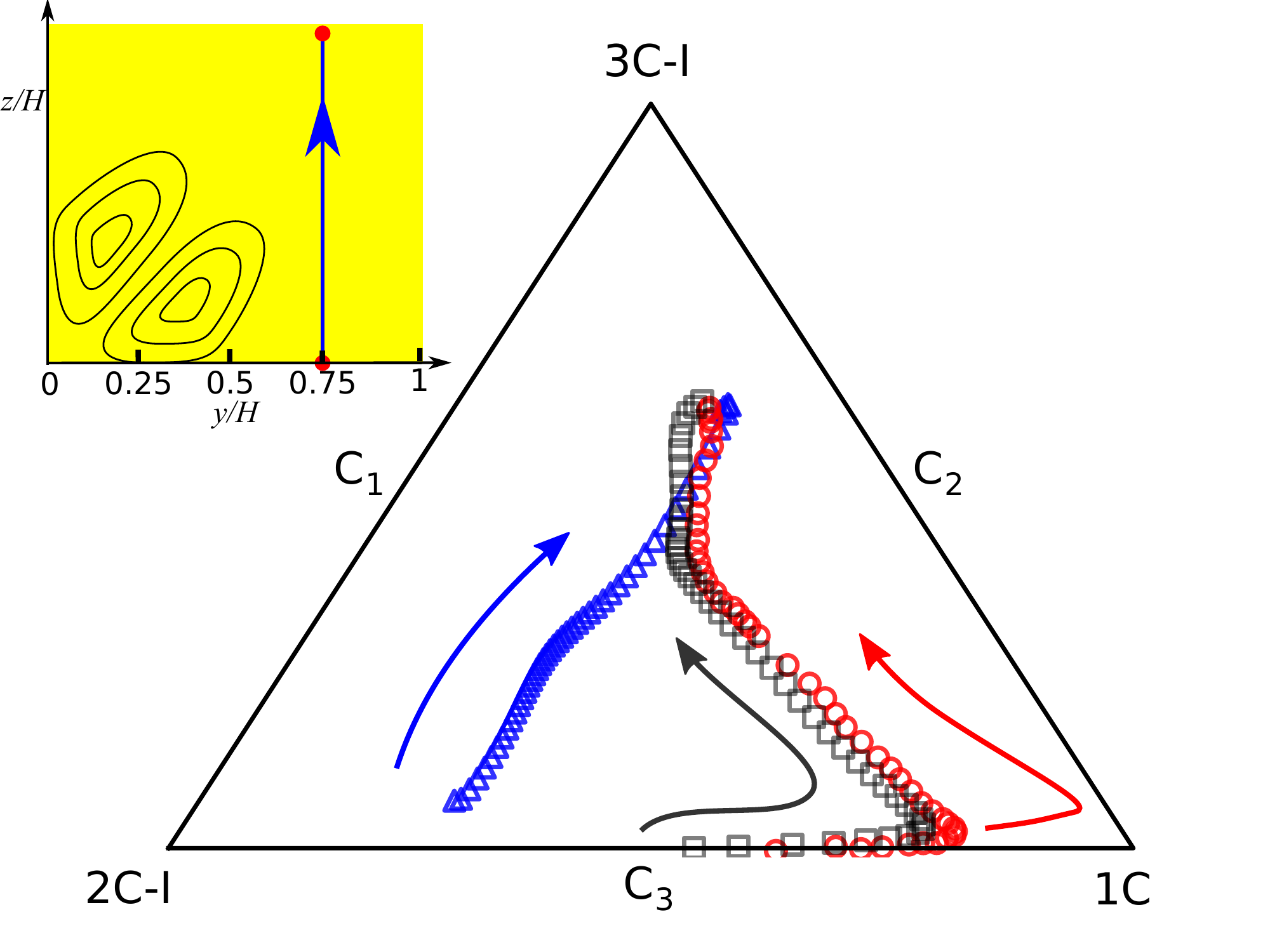}}
	\caption{Barycentric map of the predicted Reynolds stress anisotropy for the test flow 
		($Re = 3500$), learned from the training flows ($Re = 2200, 2600$, and $2900$)
		The prediction results on two streamwise locations at $y/H = 0.25$ and $0.75$ are compared with
		the corresponding baseline (RSTM) and DNS results in panels (a) and (b), 
		respectively.}
	\label{fig:bayRe}
\end{figure}

{We first investigate the prediction performance on the Reynolds stress anisotropy tensor,
	since its accuracy is important to capture the secondary flow. Figure~\ref{fig:bayRe} shows 
	PIML-corrected anisotropy in Barycentric triangle compared with baseline and DNS results. 
	The comparisons are performed on two representative lines at $y/H = 0.25$ and $0.75$ 
	on the in-plane cross section (Fig.~\ref{fig:domain_duct}b). The
	two lines are indicated in the insets on the upper left corner of each panel. The arrows denote the 
	order of sample points plotted in the triangle, which is from the bottom wall to the outer layer.
	The general trends of spatial variations of the DNS Reynolds stress anisotropies are similar on 
	both lines. That is, from the wall to the outer layer, the Reynolds stress starts from the 
	two-component limiting states (bottom edge of the triangle) toward three-component anisotropic 
	states (middle area of the triangle). This trend is captured by the baseline RSTM to some 
	extent, especially in the regions away from the wall. However, significant discrepancies still
	can be observed in the near-wall region. Very close to the wall, the DNS Reynolds stress is
	nearly the two-component limiting state. This is because the velocity fluctuations in the 
	wall-normal direction are suppressed by the blocking of the bottom wall. Moreover, before 
	approaching three-component anisotropic states, the DNS-predicted anisotropy first moves 
	toward the one-component state (1C) as away from the wall.}
{In contrast, the RANS-predicted anisotropy near the wall is closer to the 
	two-component isotropic state (2C-I), and it approaches toward the three-component anisotropic
	state directly. Therefore, in the near-wall region there are large discrepancies between the RANS 
	predicted Reynolds stress anisotropy and the DNS result, particularly in the horizontal 
	coordinate~$\xi$. By correcting the baseline RSTM results with the trained discrepancy function, 
	the predicted anisotropy of Reynolds stress is significantly improved. For both lines, the predicted 
	anisotropy (circles) agrees well with the DNS results (squares). Especially on the line 
	$y/H = 0.75$, the PIML-predicted anisotropy is almost identical to the DNS data.}

{Significant improvement of the PIML-predicted anisotropy can be seen from the
Barycentric maps shown in Fig.~\ref{fig:bayRe}. Similar improvements have also been demonstrated
in the other physical projections (TKE and orientations) of the PIML-corrected Reynolds stresses.
Therefore, it is expected that the Reynolds stress tensor components should be also improved over 
the RSTM baseline.}
\begin{figure}[htbp]
	\centering
	\includegraphics[width=0.4\textwidth]{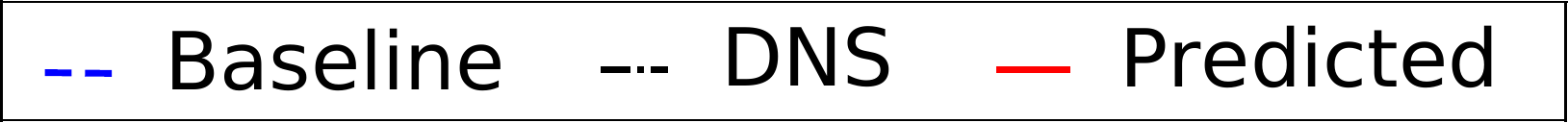}\\	
	\subfloat[$\tau_{yy}$]{\includegraphics[width=0.45\textwidth]{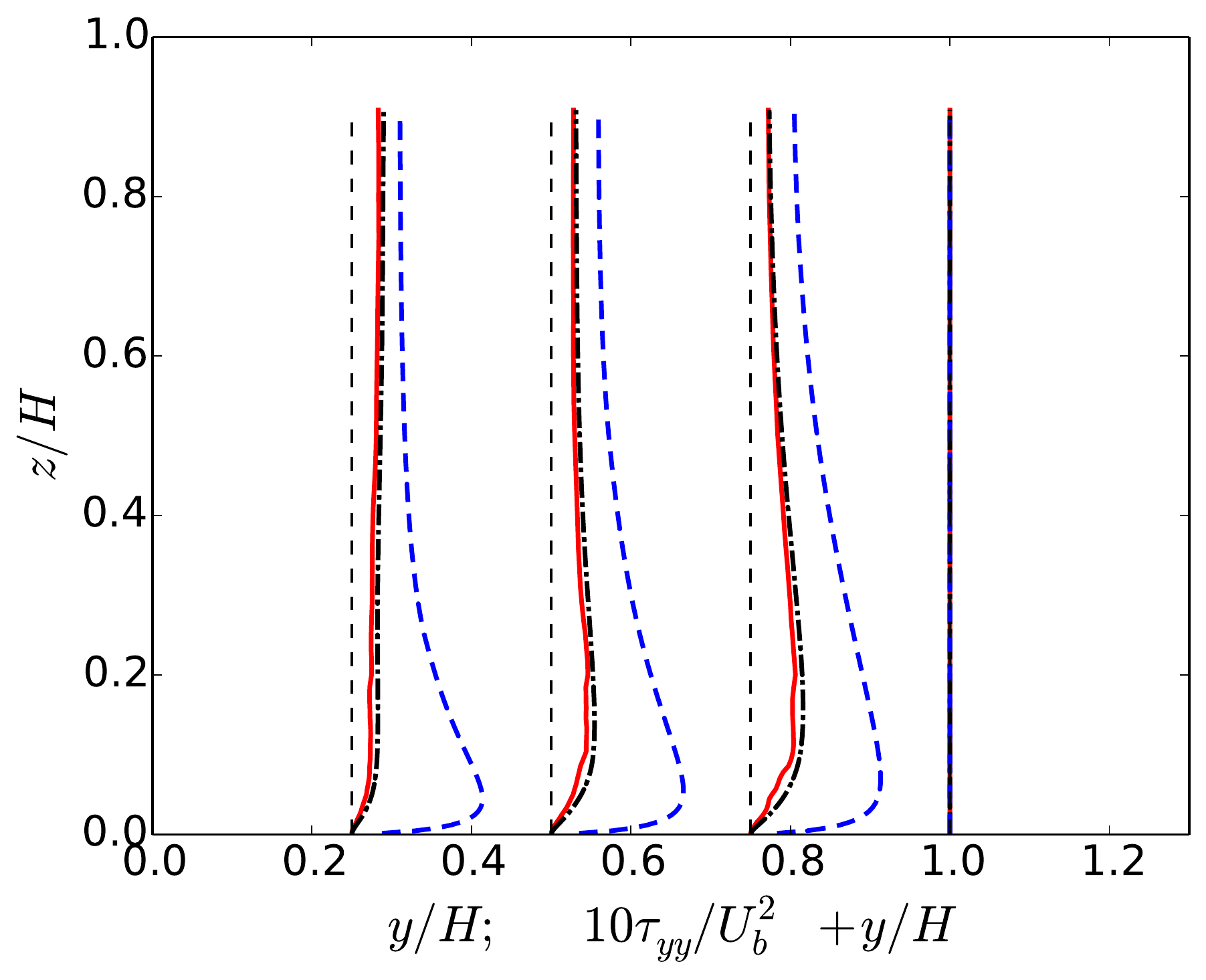}}
	\subfloat[$\tau_{zz}$]{\includegraphics[width=0.45\textwidth]{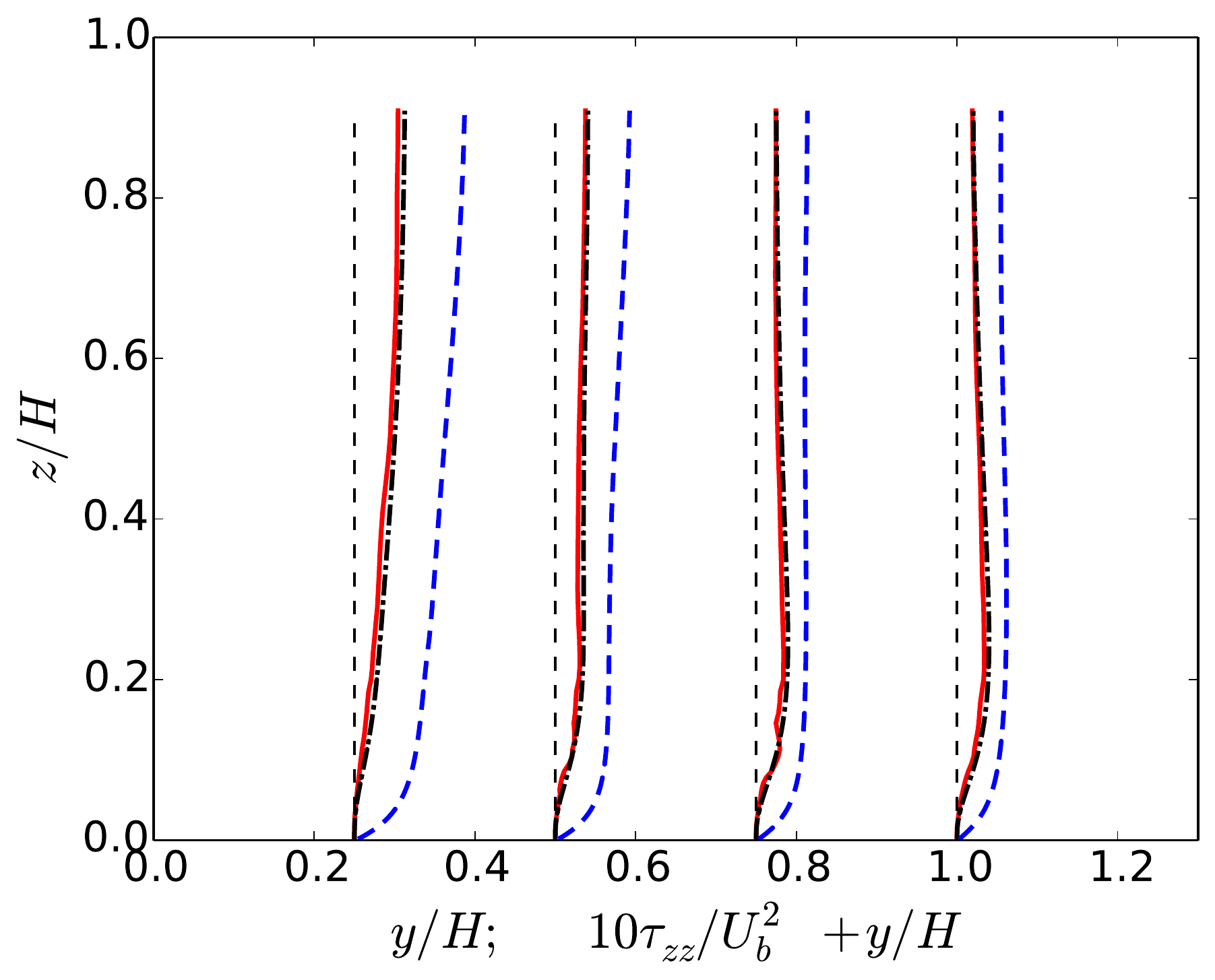}}
	\caption{{Profiles of normal components (a) $\tau_{yy}$ and (b) $\tau_{zz}$ of corrected 
		Reynolds stress with the discrepancy model. The profiles are shown at four streamwise 
		locations $y/H = 0.25, 0.5, 0.75,$ and $1$. Corresponding DNS and baseline (RSTM) 
		results are also plotted for comparison.}}
	\label{fig:tau_duct}
\end{figure}
{In the six tensor components, two normal stress components $\tau_{yy}$ and $\tau_{zz}$
	are among the most important ones to the mean velocity field since the normal stress imbalance
	($\tau_{yy} - \tau_{zz}$) is the driving force of the secondary flow~\cite{bradshaw1987turbulent}.
	Figures~\ref{fig:tau_duct}a and~\ref{fig:tau_duct}b show the profiles of normal 
	components $\tau_{yy}$ and $\tau_{zz}$ of the Reynolds stress reconstructed from the
	PIML-corrected physical projections. Corresponding baseline (RSTM) and DNS results
	are also plotted for comparison. Both $\tau_{yy}$ and $\tau_{zz}$ are overestimated
	by the RSTM over the entire domain. The discrepancy of the RSTM-predicted $\tau_{yy}$ 
	is large near the wall and decreases when moving away from the wall. In contrast, $\tau_{zz}$ is 
	significantly overestimated far from the wall but the discrepancy decreases toward the
	wall. As a results, the RSTM-predicted normal stress imbalance is pronouncedly inaccurate,
	which leads to unreliable secondary mean flow motion. As expected, the PIML predictions
	nearly overlap with the DNS results for both $\tau_{yy}$ and $\tau_{zz}$ and show
	considerable improvements over the RSTM baseline. In fact, the improvements are observed
	in all the tensor components, which are omitted here for brevity. The results shown above 
	demonstrate excellent performance of the proposed PIML framework by using RSTM
	as the baseline.}

\subsection{Turbulent Flows With Massive Separations}
\label{sec:result:separation}

\subsubsection{Case Setup}
\label{sec:result:separation:setup}
{The turbulent flow in a channel with periodic hills is another challenging case for RANS
models due to the massive flow separations in leeward of the hill. Here, we examine two 
training scenarios with increasing difficulty levels.} In the first scenario the training flows have 
the same geometry as the test (prediction) flow but are different in Reynolds numbers. In the 
second scenario the training flows differ from the prediction case not only in Reynolds numbers 
but also in geometry.
\begin{table}[htbp]
	\centering
	\caption{
		Database of training flows to predict flow past periodic hills at $Re = 10595$. The
		Reynolds numbers are defined based on the bulk velocity $U_b$ at the narrowest
		cross-section in the flow and the crest/step height $H$.
	}
	
	\label{tab:database}
	\begin{tabular}{P{2.5cm} | P{8.0cm} | P{4.5cm}  }	
		\hline
		Training flow scenario & Training flow \& symbol & High fidelity data\\ 
		\hline
		\multirow{ 2}{*}{Scenario \RN{1}} & Periodic hills, Re = 1400 (PH1400) & DNS by Breuer et al.~\cite{breuer2009flow} \\
		& Periodic hills, Re = 5600 (PH5600) & DNS by Breuer et al.~\cite{breuer2009flow} \\
		\hline
		\multirow{ 2}{*}{Scenario \RN{2}} & Wavy channel, Re = 360 (WC360) & DNS by Maa{\ss} et al.~\cite{maass1996direct} \\
		& Curved backward facing step, Re = 13200 (CS13200)& LES by Bentaleb et al.~\cite{bentaleb2012large} \\
		\hline								
	\end{tabular}
\end{table}

Four training flows with DNS/LES data to build random forest regressors are summarized in
Table~\ref{tab:database}. In the first scenario two flows PH1400 and PH5600 are used for training,
both of which are flows over periodic hills (same in geometry) at $Re=1400$ and $Re=5600$ (different
in Reynolds numbers), respectively. For the second scenario, the training data are obtained from two
different flows: one in a channel with a wavy bottom wall at $Re = 360$ and one over a curved
backward facing step at $Re = 13200$, indicated as flows WC360 and CS13200, respectively.

\begin{figure}[htbp]
	\centering \subfloat[Periodic hills (Re = 1400 and 5600, results from latter are shown)] {\includegraphics[width=0.8\textwidth]{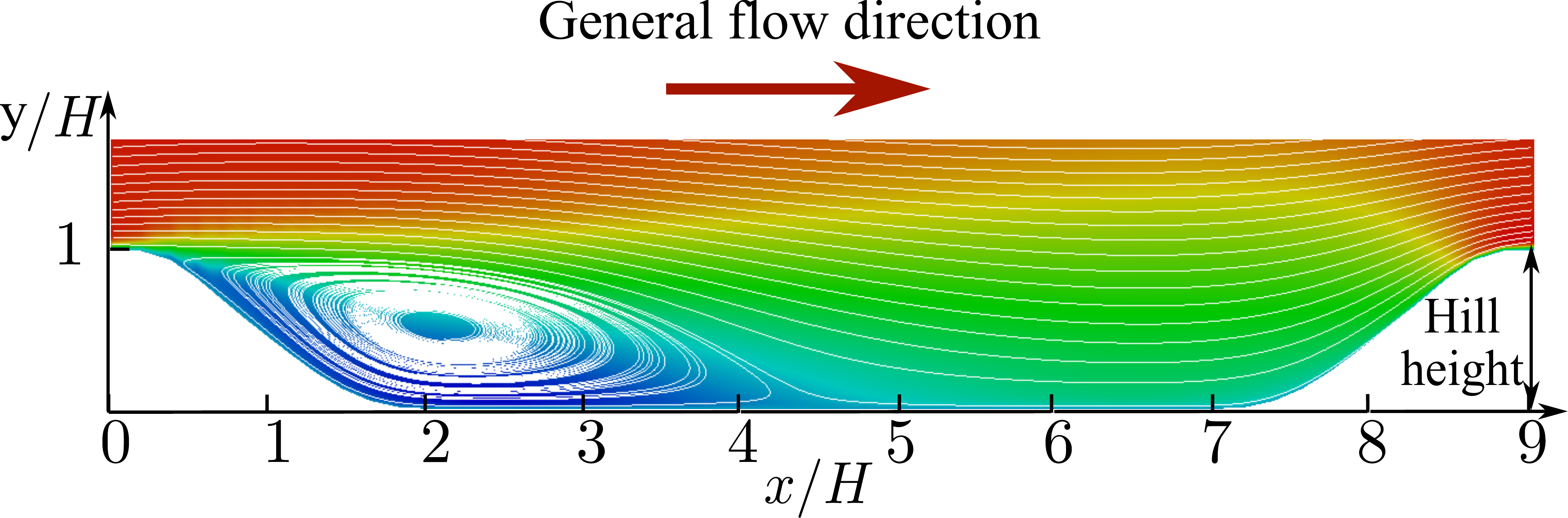}}\\
	\subfloat[Wavy channel (Re = 360)] {\includegraphics[width=0.8\textwidth]{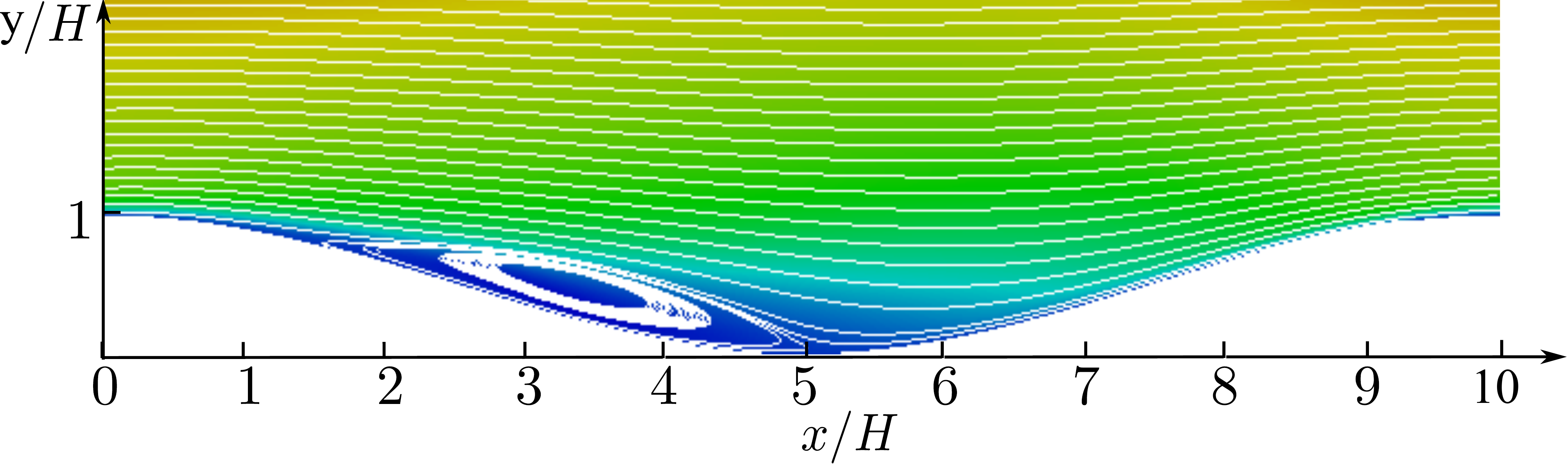}}\\
	\subfloat[Curved backward facing step (Re = 13200)] {\includegraphics[width=0.8\textwidth]{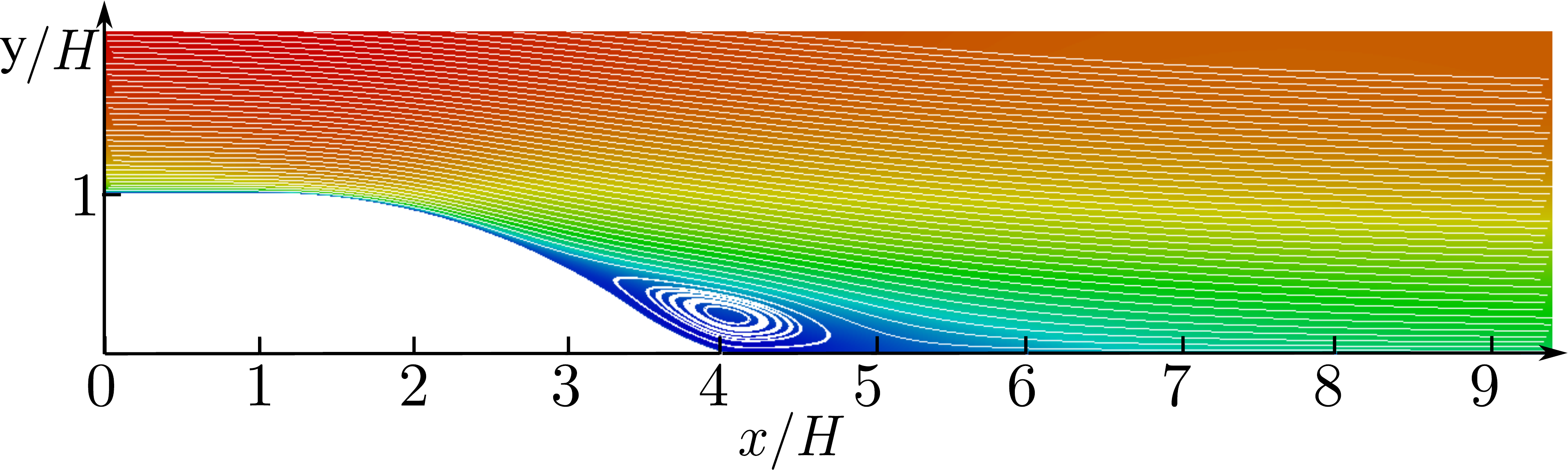}}
	\caption{Computational domain and velocity field of each case in the training flow database. 
		The velocity contours and streamlines are obtained from the baseline RANS simulations. 
		The dimensions of each case are normalized with the respective hill heights $H$. 
		Note that periodic boundary conditions are applied on the flows in panels (a) and (b)
		in steamwise direction, but not for the flow in panel (c). }
	\label{fig:flowGeo}
\end{figure}
A schematic of the flow geometry and RANS-predicted velocity contour for each case are 
presented in Fig.~\ref{fig:flowGeo}. The dimensions of each case are normalized with the respective
hill heights~$H$. Although the geometries of the training flows are different, all three flows share a similar 
characteristic as the test flow, i.e., separation on the leeward side of the hill or step. However, the separation 
bubbles are different in sizes and shapes. The flow over periodic hills has a stronger separation compared 
to the other two due to the steeper slope of the hill. Relatively mild separation can be observed in the flow 
over the wavy channel. For all cases, both high-fidelity data and RANS-predicted results are available. 
The high-fidelity data are obtained from DNS or 
resolved LES simulations, which have been reported in literature (see references in Table.~\ref{tab:database}). 
The DNS data for flows PH1400 and PH5600 are only available on vertical lines at eight streamwise locations
$x/H = 1, 2, \cdots, 8$. On the other hand, full-field high-fidelity results are available for flows WC360 
and CS13200, but only the lower part of the channel is adequately resolved. Since the separated flow is 
of interest in this study, only the data in the separation region (i.e., region below $y/H = 1.2$) are included.

{In this test, the performance of the proposed PIML framework is evaluated on standard RANS models. 
Specifically, the baseline RANS predictions are obtained by using the two-equation Launder-Sharma
$k$--$\varepsilon$ model~\cite{launder1974application}.} 
The reason for choosing standard turbulence models here is because of two considerations. 
First, the standard RANS models are the dominant tools for industrial CFD applications, 
while other sophisticated RANS models have been rarely used. Therefore, it is 
more significant to improve the widely used standard RANS models. Second, we understand 
that improvement of the Reynolds stresses starting from a standard RANS model is 
challenging. Nonetheless, this challenging scenario also can better explore the capability of 
machine learning approach. 
\subsubsection{Prediction Results}
{The functional forms of discrepancies in the six physical projections of Reynolds stress are
	learned from the training flows as mentioned in Section~\ref{sec:result:separation:setup} 
	and are used to correct the RANS-predicted Reynolds stress field of the test flow (PH10595). 
	However, since the baseline RANS model used in this case is the standard eddy viscosity model, 
	the Reynolds stress anisotropy cannot be accurately predicted. Therefore, the baseline 
	RANS-predicted anisotropy is unphysical and is significantly different from the DNS result 
	(see Fig.~\ref{fig:bary}). Nonetheless, after the correction by using the discrepancy function learned from 
	the training flows, the anisotropy of the test flow shows an excellent agreement with the DNS 
	results~\cite{xiao-perspect}. The improvements are observed in the both training scenarios I and II, 
	demonstrating that the discrepancy function even in the standard RANS-predicted anisotropy 
	does exist and can be learned from the closely related flows based on the mean flow 
	features $\mathbf{q}$. As mentioned above, in the periodic-hill flow, the correctness of 
	Reynolds stress anisotropy is of little consequence to the prediction of the mean velocity, and the 
	correct shear stress component and magnitude of the Reynolds stress are most important 
	to obtain an accurate mean flow field. Therefore, the anisotropy prediction results are omitted 
	here, and only the turbulence kinetic energy (TKE) and shear stress component of the PIML-corrected 
	Reynolds stress are presented and discussed in detail.}

\begin{figure}[htbp]
	\centering
	\includegraphics[width=0.8\textwidth]{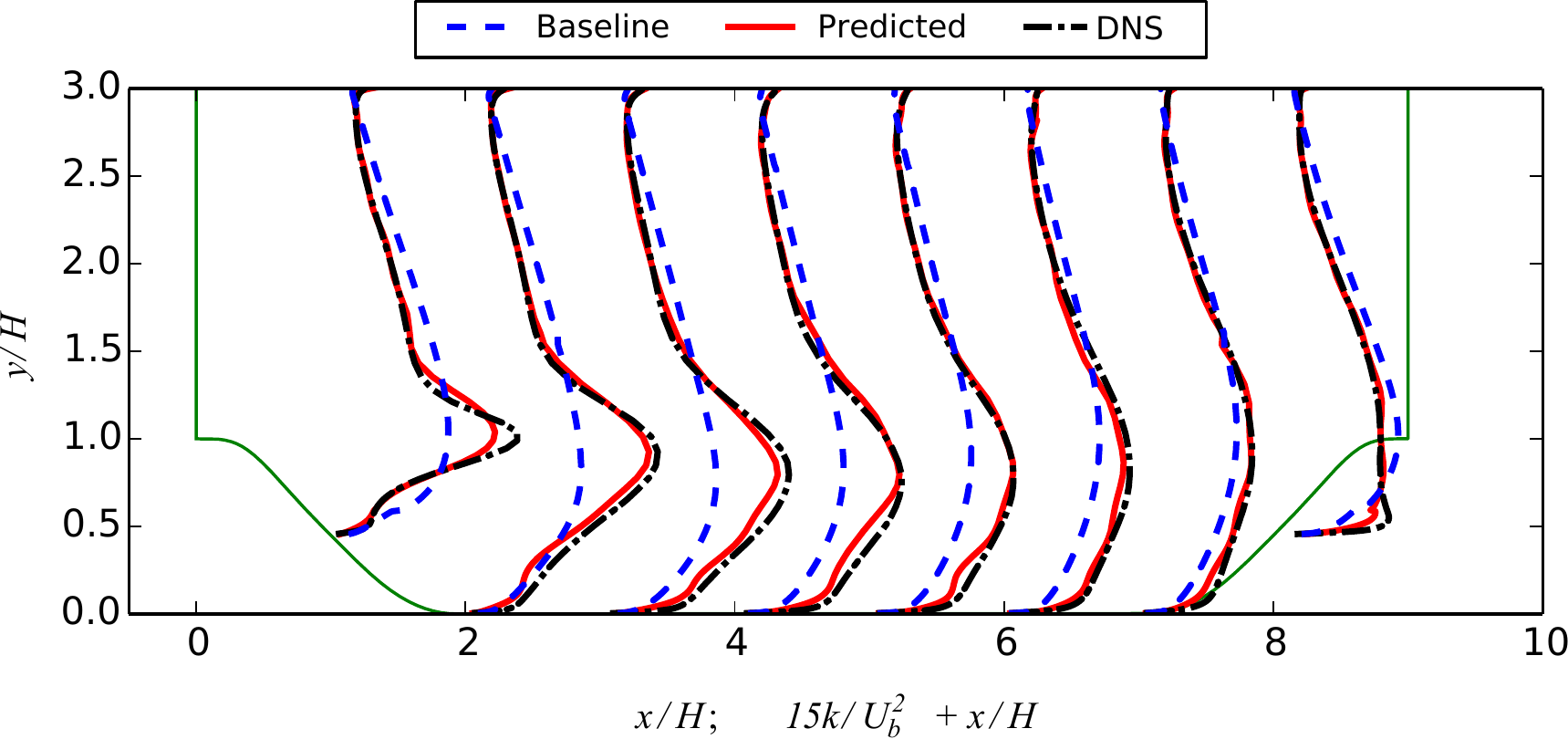}
	\caption{Magnitude (turbulence kinetic energy) of the corrected Reynolds stress for the test flow 
		(PH10595) learned from cases with same geometry but at different Reynold numbers (PH1400 and PH5600). 
		The profiles are shown at eight streamwise locations $x/H = 1, 2, \cdots, 8$. 
		Corresponding baseline and DNS results are also plotted for comparison. 
		The hill profile is vertically exaggerated by a factor of 1.3.}
	\label{fig:KRe}
\end{figure}

{The comparison of the TKE profiles of the baseline, DNS, and PIML-predicted results in the 
	training scenario I are shown in Fig.~\ref{fig:KRe}. The TKE predicted by the baseline RANS 
	model has notable discrepancies compared to the DNS result, particularly in the region with 
	non-parallel free shear flow ($y/H = 0.8$ to $1.5$).} The poor performance of RANS model 
	in such region is typical in these flows~\cite{xiao-mfu}. The RANS model underestimates the 
	turbulence intensity along the free shear at $y/H = 1$, especially near the leeward side of the 
	hill ($x/H = 1$ to $2$). In the upper channel ($y/H = 1.5$ to $2.5$), the DNS TKE is slightly 
	smaller than the baseline RANS prediction. The profiles of TKE corrected by the PIML-predicted 
	discrepancy $\Delta \log k$ are significantly improved. The peaks along the streamwise free 
	shear in the DNS profiles are well captured in the corrected results with the random forest 
	prediction. It can be seen that the predicted TKE profiles (solid lines) nearly overlap with the 
	DNS results (dashed lines). This clearly indicates that the TKE discrepancies can be learned 
	from the data of the training flows. 

\begin{figure}[htbp]
	\centering
	\includegraphics[width=0.8\textwidth]{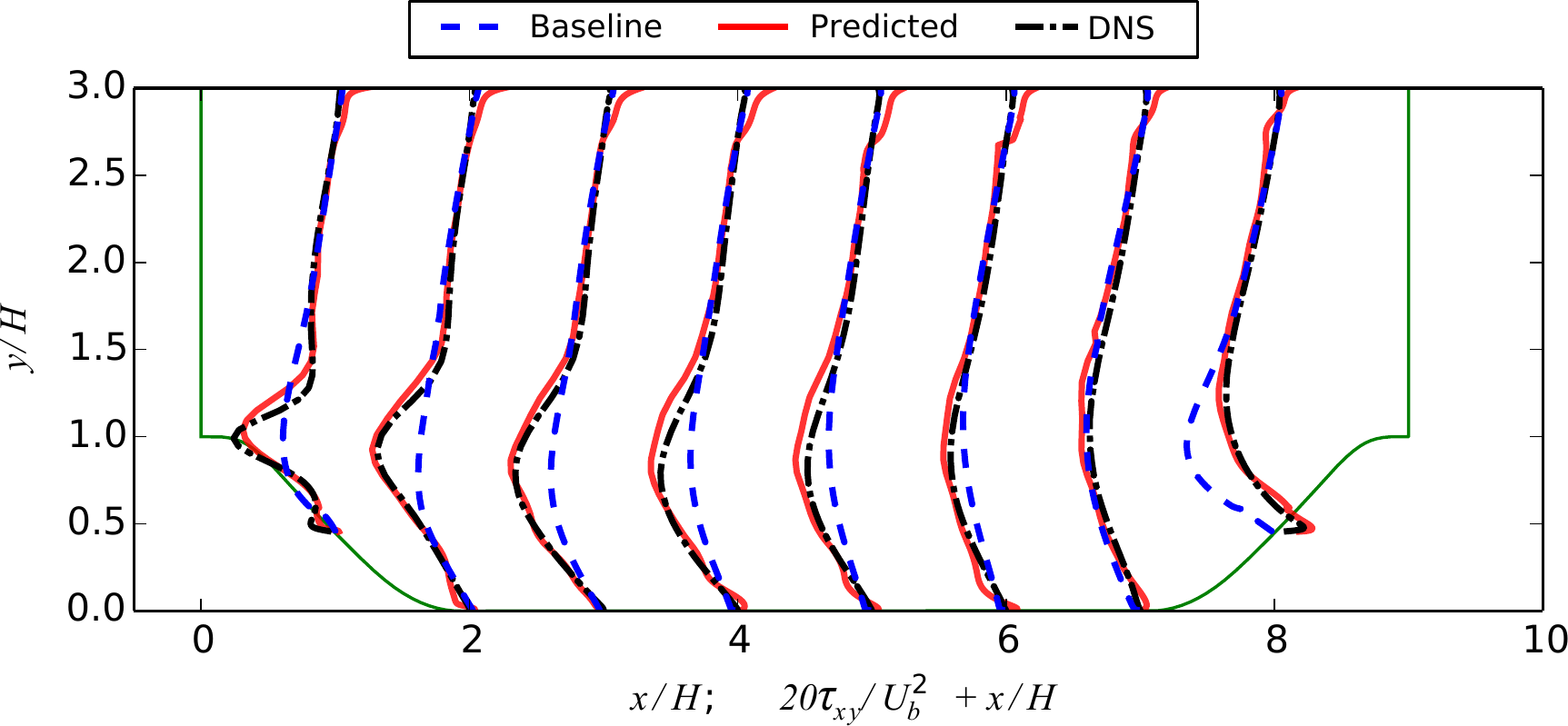}
	\caption{Predicted turbulence shear stress for the test flow (PH10595) learned from the flows with 
		the same geometry but at different Reynold numbers (PH1400 and PH5600). The profiles are 
		shown at eight streamwise locations $x/H = 1, 2, \cdots, 8$. Corresponding baseline and 
		DNS results are also plotted for comparison. The hill profile is vertically exaggerated by a 
		factor of 1.3.}
	\label{fig:TauRe}
\end{figure}

It is also of interest to investigate the tensor components $\tau_{ij}$ of Reynolds stress, which
are more relevant for predicting velocities and other QoIs of the flow fields. For the plane shear flows, 
the turbulent shear stress $\tau_{xy}$ is important to predict the velocity field. Figure~\ref{fig:TauRe} 
compares the turbulence shear component~$\tau_{xy}$ of predicted Reynolds stress with the DNS. 
As expected, significant improvements are observed compared to the baseline results, which underestimate 
the peak of $\tau_{xy}$ on the leeward hill side but overestimate it on the windward hill side. As shown in Fig.~\ref{fig:TauRe}, the profiles of predicted $\tau_{xy}$ agree well with the DNS results. 

The results above demonstrate that the discrepancy function of Reynolds stress in its physical 
projections (i.e., magnitude, shape, and orientation) trained from the flows at Re = 2800 and 5600 
can be used to predict the Reynolds stress field of the flow at Re = 10595. Significant improvements
are observed in the predicted Reynolds stress compared to the baseline RANS results. Although 
in this scenario the training and test flows are quite similar (with the same geometry), and 
the success of extrapolation has been demonstrated in physical space by Wu et al.~\cite{wu2016bayesian}, 
it should not be taken for granted that the accurate prediction is also guaranteed in feature space. 
Since the regressions are performed in the ten-dimensional feature space and there is no 
direct reference to the physical coordinate, the success is not trivially expected \emph{a priori}.

We investigate a more challenging scenario where the training flows have different geometries from 
the prediction case. This scenario is also more realistic in the context of using RANS simulation to 
support engineering design and analysis. Specifically, the data are more likely to be available for a 
few flows with specific Reynolds numbers and geometries, but predictions are needed for the 
similar flows yet at different Reynolds numbers and with modified geometries.

\begin{figure}[htbp]
	\centering
	\includegraphics[width=0.9\textwidth]{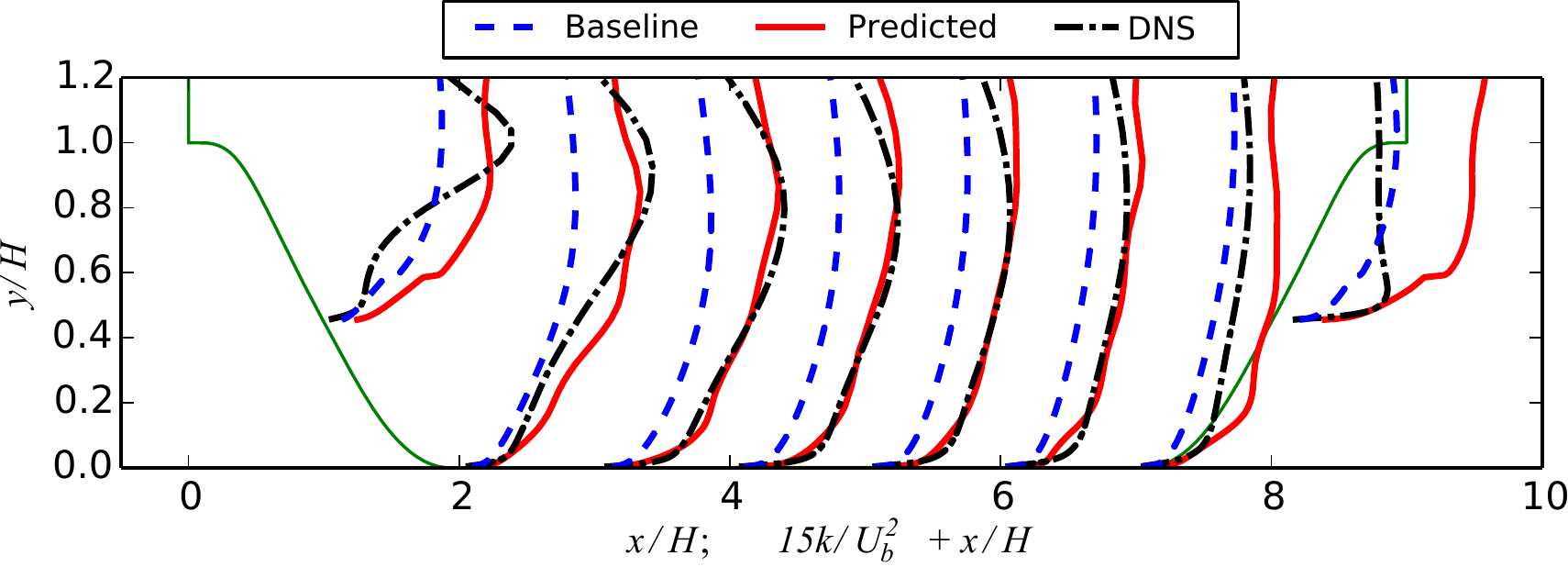}
	\caption{
		Magnitude (turbulence kinetic energy) of the corrected Reynolds stress for the test flow 
		(PH10595) learned from cases with different geometries and at different Reynold numbers 
		(WC360 and CS13200). The profiles are shown at eight streamwise locations $x/H = 1, 2, \cdots, 8$. 
		Corresponding baseline and DNS results are also plotted for comparison. 
		The hill profile is vertically exaggerated by a factor of 2.4.
	}
	\label{fig:KGeo}
\end{figure}

{The comparison of the TKE profiles on eight lines is shown in Fig.~\ref{fig:KGeo}. Note that 
	only the domain below $y/H = 1.2$ is investigated due to the lack of reliable high-fidelity training 
	data in the upper channel region. This inadequacy of data quality can be exacerbated when the 
	Reynolds stress is decomposed to its physical projections. Moreover, the flow separation is the 
	phenomenon of concern in this study, and thus we only focus on the recirculation region.}
 {In Fig.~\ref{fig:KGeo}, the random forest predicted TKE (solid lines) is 
	significantly improved over the baseline results (dotted lines) and better agrees with the DNS profiles (dash-dotted lines).} The agreement is particularly good in the region from the center of 
recirculation bubble ($x/H =2$) to the beginning of flow contraction ($x/H = 6$). {Nonetheless, 
	the PIML-predicted TKE does not show any improvement and even deteriorates compared to the baseline 
	results near the windward side of the hill ($x/H > 7$), where the flow starts to be contracted. }
 As shown in Fig.~\ref{fig:KGeo}, the predicted TKE is markedly overestimated at $x/H = 8$. {This is because 
	the flow features in the contraction region ($x/H > 7$) are not supported in the training set, since 
	the contracted flow does not exist in the training flow CS13200 (Fig.~\ref{fig:flowGeo}c) and is much 
	weaker in the training flow WC360 (Fig.~\ref{fig:flowGeo}b) due to the mild slope of wavy bottom in 
	this geometry.}

\begin{figure}[htbp]
	\centering
	\includegraphics[width=0.9\textwidth]{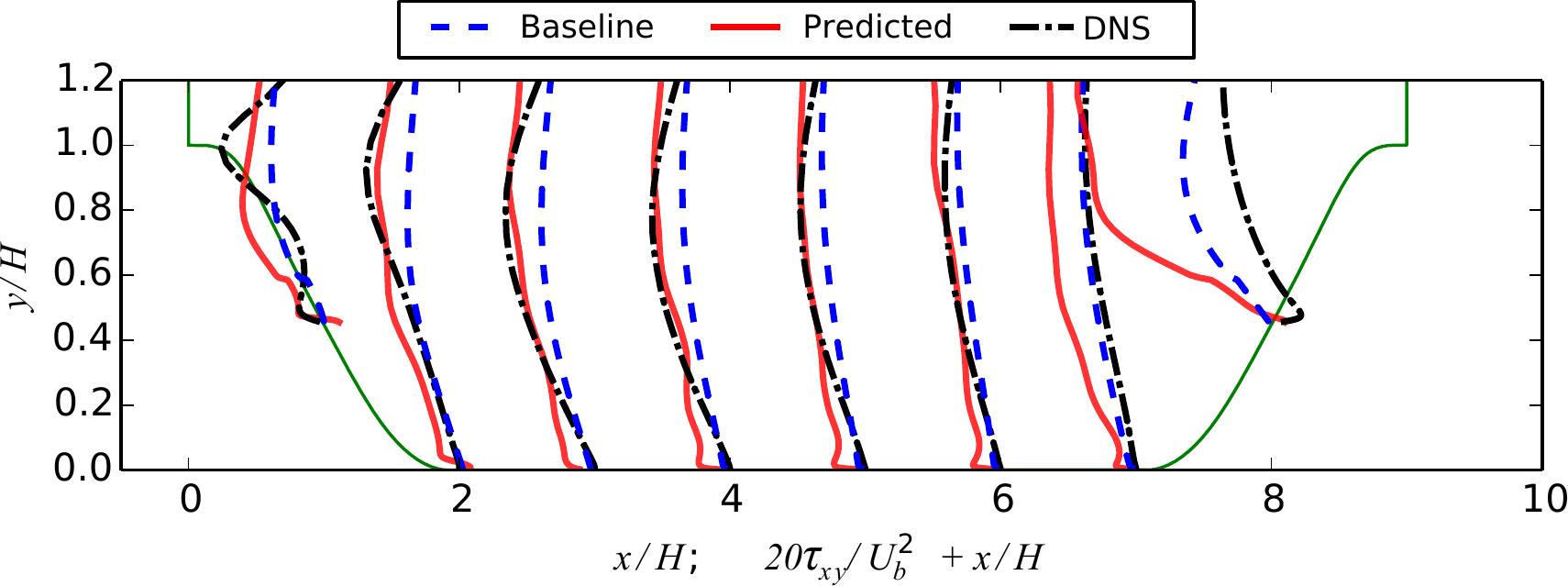}
	\caption{
		Predicted turbulence shear stress for the test flow (PH10595) learned from the flows with different 
		geometries and at different Reynold numbers (WC360 and CS13200).  The profiles are 
		shown at eight streamwise locations $x/H = 1, 2, \cdots, 8$. Corresponding baseline and 
		DNS results are also plotted for comparison. The hill profile is vertically exaggerated by a 
		factor of 2.4.	
	}
	\label{fig:TauGeo}
\end{figure}

Finally, we compare the predicted turbulent shear stress $\tau_{xy}$ with the DNS profiles in
Fig.~\ref{fig:TauGeo}. {Similar to the results of TKE, the PIML-predicted turbulent shear stress 
$\tau_{xy}$ shows notable improvements in the recirculation region.}  However, deterioration occurs in 
the flow contraction region. At $x/H = 7$ and $8$, the magnitudes of turbulent shear stresses are 
overestimated with the correction based on the predicted discrepancies. This is consistent with 
the results observed in physical projections of Reynolds stress. Such a small region with abnormal 
Reynolds stress corrections (artificial peaks or bumps) can introduce large errors to the velocity 
predictions. 

In general, the physical projections (i.e., magnitude, shape, and orientation) of Reynolds stress 
corrected by random forest predicted discrepancies are still significantly improved with the training 
flows in different geometries (WC360 and CS13200). The Reynolds stress is markedly improved in the 
separated flow region, but not in the contracted flow region. This is because the features in training 
flows cannot well support the predicted flow, and thus more extrapolations are expected. Although 
the improvement is less significant compared to that in the scenario \RN{1}, the random forest predictions 
in this more realistic scenario are still satisfactory, demonstrating the merits of the proposed PIML framework.

\section{Discussion}

\subsection{Feature Importance and Insight for Turbulence Modeling}
\label{sec:important}

In addition to the predictive capability of the regression model, it is also important to interpret 
the functional relation between the mean flow features and the discrepancies of the RANS modeled 
Reynolds stresses. For example, it is useful to find the most important features to
the Reynolds stress discrepancy in each of its physical projection (i.e., magnitude $k$, 
shape $\xi, \eta$, and orientation $\varphi_i$), and how each of these features impacts the
regression response. 
\begin{figure}[htbp]
	\centering 
	\subfloat[$\Delta \eta$]
	{\includegraphics[width=0.45\textwidth]{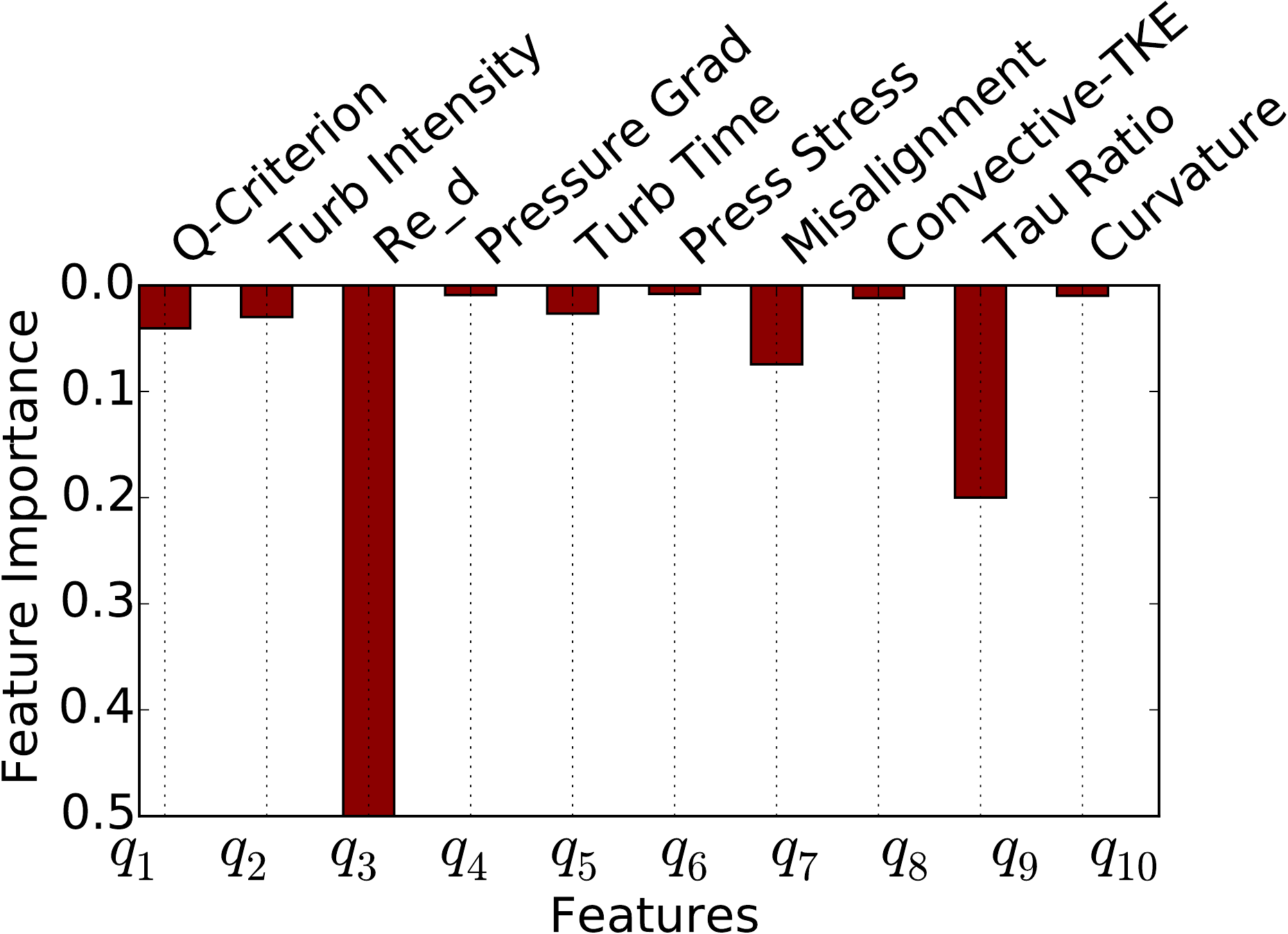}} \hfil
	\subfloat[$\Delta\log k$]
	{\includegraphics[width=0.45\textwidth]{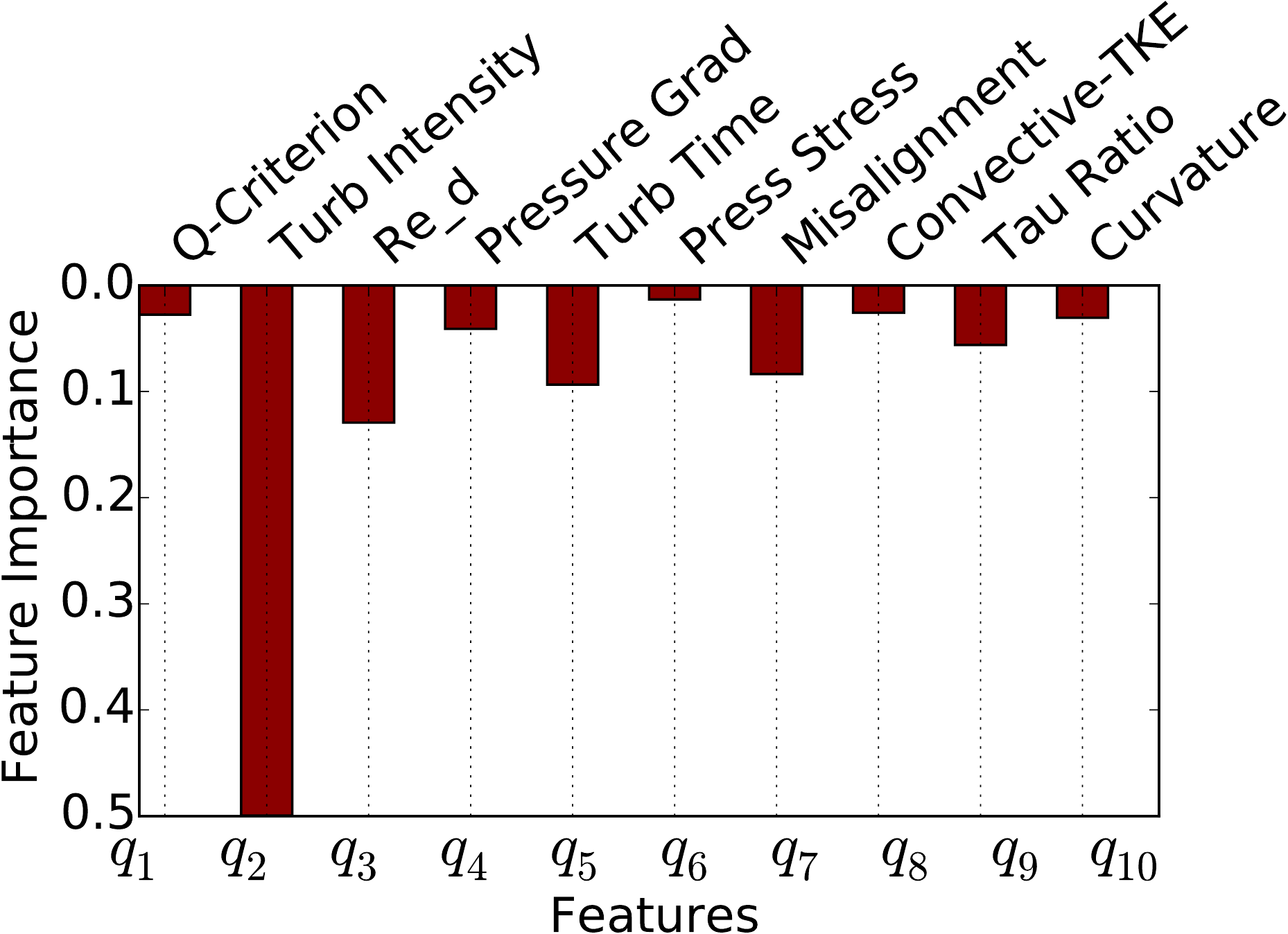}}
	\caption{Feature importance of random forests regressors (a) for $\Delta\eta$ and 
		(b) for $\Delta \log k$ for scenario \RN{1} (i.e., training flows in the same geometry,
		see Table.~\ref{tab:database}). The features $q_i$ ($i = 1$ to 10) are 
		denoted by their respective abbreviations. Turb Intensity denotes the turbulence 
		intensity (feature $q_2$), and Re\_d is the wall distance based Reynolds 
		number (feature $q_3$). For the full name list of features, see Table~\ref{tab:feature}.}
	\label{fig:import}
\end{figure}
Identification of such correlation or causal relationship enables modelers to improve
the RANS turbulence models. The random forest regressor used in the proposed PIML 
framework can also shed light on this issue by calculating the feature importance, which is a
measure to evaluate the relative importance of a feature variable for predicting
response variables~\cite{breiman2001random}. The bar plots of the importances of feature 
vector $\mathbf{q}$ with respect to the discrepancies $\Delta \eta$ and $\Delta\log k$ 
are shown in Figs.~\ref{fig:import}a and~\ref{fig:import}b, respectively. For discrepancy 
$\Delta \eta$ in the anisotropy, feature $q_3$ (i.e., wall-distance based Reynolds 
number $Re_d$) is the most important one. As discussed in Section~\ref{sec:feature}, 
$Re_d$ is the wall distance normalized by the approximate viscous unit. Therefore, 
the result of feature importance is consistent with the PIML prediction, which have shown 
that the discrepancy $\Delta \eta$ is notably dependent on the distance away from 
the wall~\cite{xiao-perspect}. Figure~\ref{fig:import}b shows that the 
most important feature for predicting discrepancy $\Delta\log k$ of turbulence kinetic energy 
is feature $q_2$, turbulence intensity. 

It is demonstrated that random forest used in the proposed framework can interpret 
the relationship between the features and the response to a certain extent, 
although the feature importance has its limitation due to bias introduced under certain 
conditions~\cite{dobra2001bias, strobl2007bias}. In the machine learning
community, improving interpretability of random forest is an active research topic,
e.g., several improvements of the importance measure have been 
proposed~\cite{strobl2007bias,sandri2012bias,altmann2010permutation}. 
Moreover, in addition to calculating the feature importance, it is also helpful to examine the 
interactions among features, which have important implications for the interpretation
of the regression models. As the base learner in random forest is a decision tree, which
can capture the feature interactions, it is possible to further investigate the interacting 
relationship among mean flow variables. Breiman et al. have studied feature 
interaction in random forest method~\cite{breiman2001random}, but more research 
is still ongoing. A better understanding of the physics behind the regression model 
for Reynolds stress discrepancies has a profound implication to RANS turbulence 
modeling. Therefore, to explore the correlation or causal relationship between the 
mean flow features and the discrepancies of RANS modeled Reynolds stress is an 
important and promising extension of the proposed framework.

\subsection{Success and Limitation of The Current Framework}
\label{sec:limitation}
{   The objective of the proposed framework is to improve the baseline RANS-predicted
	Reynolds stresses of a flow where high-fidelity (e.g., DNS, LES, experimental) data are not 
	available. The main novelty lies in using machine learning techniques to find the functional
	forms of Reynolds stress discrepancies with respect to mean flow features by learning from the existing
	offline database of the closely related flows. Numerical simulation results have demonstrated 
	the feasibility and merits of the framework. Moreover, the excellent performance of the PIML
	predicted Reynolds stress in not only the anisotropy but also in the TKE and turbulent shear 
	stress shows the fact that Reynolds stress discrepancies can be extrapolated even to complex 
	flows sharing similar characteristics. This finding is noteworthy by itself.} 
		
	The improvement of the RANS-predicted Reynolds stress is considered a viable and 
	promising path toward obtaining better predictions of velocities and other quantities of 
	interest. However, due to a few limitations of the current framework, the improvement of
	the propagated velocities from the corrected Reynolds stress field can not be guaranteed.
	A small region with abnormal Reynolds stress corrections (e.g., non-smoothness or artificial peaks) 
	can introduce large errors to the velocity predictions. For example, the small wave-number variations 
	in Reynolds stresses are visible in Fig.~\ref{fig:TauGeo}. These fluctuations, despite being small in 
	amplitude, can lead to abnormal behaviors in the divergence term and thus in the predicted 
	velocities. These abnormal predictions of Reynolds stress corrections can be caused by several 
	factors. First, the features in certain regions of the prediction flow may not be well supported in 
	the training flows, e.g. the contraction region of periodic-hill flow mentioned in 
	Sec.~\ref{sec:result:separation}. Second, the random forest regression used here only
	provides pointwise estimations but cannot consider the spatial information of the Reynolds
	stress field. Therefore, the smoothness of the prediction cannot be guaranteed. Finally, 
	although the input feature space is constructed based on physical reasoning, it is possible that the
	input features are not rich enough, and thus the randomness in the ensemble of the trained 
	decision trees is significant.

\section{Conclusion}

{In this work, we proposed a physics-informed machine learning approach to reconstruct Reynolds stresses modeling discrepancies by utilizing DNS databases of training flows sharing similar characteristics as the
flow to be predicted.} For this purpose, we formulated discrepancy of Reynolds stresses (or more
precisely its magnitude and the shape and orientation of the anisotropy) as target functions of mean
flow features and used modern machine learning techniques based on random forest regression to learn
the functions. The obtained functions are then used to predict Reynolds stress discrepancies in new
flows. {To evaluate the performance of the proposed approach, the method is tested by two 
	classes of flows: (1) fully developed turbulent flows in a square duct at various Reynolds numbers 
	and (2) flows with massive separations. In the separated flows, two training flow scenarios of increasing 
	difficulties are considered:}
In the less challenging scenario, data from two flows in the same periodic hill geometry at lower Reynolds 
numbers ($Re=2800$ and 5600) are used for training. In a more challenging scenario, the training data come
from separated flows in different geometries (wavy channel and curved backward facing step). In all
test cases the corrected Reynolds stresses are significantly improved compared to the baseline RANS
predictions, demonstrating the merits of the proposed approach. In the scenario, where the training 
flows and the prediction flow have different geometries, the improvement is not as drastic as in the 
the scenario with the same geometry. This is expected since the prediction involves more extrapolations 
in the feature space for this more challenging scenario. In other words, compared to the first scenario
where the training and prediction flows have identical geometry, the prediction flow is less
``similar'' to the training flows in this scenario. The extent to which the training and prediction
flows are ``similar'' to each other can be assessed \textit{a priori} based on their respective RANS
predicted mean flow field, and methods for such assessment are presented in companion publications~\cite{mfu9,wu-tsnet}.

As the inaccuracy in modeled Reynolds stresses is the dominant source of model-form uncertainty in
RANS simulations, the proposed method for improving RANS-predicted Reynolds stresses is an important
step towards the goal of enabling predictive capabilities of RANS models. Moreover, the random
forests regression technique adopted in this work can provide physical insights regarding the
relative importance of mean flow features that contributed to the discrepancies in the RANS
predicted Reynolds stresses. This information can be used to assist future model development in that
developers can devise models that are aware of and correctly respond to these flow features.
However, a number of challenges need to be tackled before the improved Reynolds stresses can be used
to predict more accurate quantities of interests that are needed in engineering design (e.g., draft
and lift coefficients). This topic will be investigated in future research.

\section*{Acknowledgment}

We thank Dr. Julia Ling of Sandia National Laboratories and Dr. Eric G. Paterson of Virginia Tech
for helpful discussions during this work. {We also thank the anonymous reviewers for
their comments, which helped improving the quality and clarity of the manuscript. }





\end{document}